\documentclass[10pt,preprint]{aastex}
\usepackage{psboxit}
\usepackage{lscape}

\PScommands

\newcommand{\vagcdir}{{\tt \$VAGC\_REDUX}}
\newcommand{\lssdir}{{\tt \$LSS\_REDUX}}
\newcommand{\Sersic}{S\'ersic}

\newcommand{\vv}[1]{{\bf #1}}

\def\simless{\mathbin{\lower 3pt\hbox
	{$\,\rlap{\raise 5pt\hbox{$\char'074$}}\mathchar"7218\,$}}} 
\def\simgreat{\mathbin{\lower 3pt\hbox
	{$\,\rlap{\raise 5pt\hbox{$\char'076$}}\mathchar"7218\,$}}} 

\newcommand{\Vmax}{\ensuremath{V_\mmax}}
\newcommand{\mmax}{\ensuremath{\mathrm{max}}}
\newcommand{\mmin}{\ensuremath{\mathrm{min}}}
\newcommand{\minmax}{\ensuremath{\mathrm{\left\{^{min}_{max}\right\}}}}

\newcounter{thefigs}
\newcommand{\fignum}{\arabic{thefigs}}

\newcounter{thetabs}

\newcounter{address}

\slugcomment{To be submitted to \aj}

\shortauthors{Blanton {\it et al.} (2004)}
\shorttitle{SDSS Galaxy Catalog}

\begin{document}
 
\title{NYU-VAGC: a galaxy catalog based on new public surveys\altaffilmark{1}}

\author{
Michael R. Blanton\altaffilmark{\ref{NYU}},
David J. Schlegel\altaffilmark{\ref{Princeton}}, 
Michael A. Strauss\altaffilmark{\ref{Princeton}},
J. Brinkmann\altaffilmark{\ref{APO}},
Douglas Finkbeiner\altaffilmark{\ref{Princeton}},
Masataka Fukugita\altaffilmark{\ref{CosmicRay}},
James E. Gunn\altaffilmark{\ref{Princeton}},
David W. Hogg\altaffilmark{\ref{NYU}}, 
\v{Z}eljko Ivezi\'{c}\altaffilmark{\ref{UW}},
G. R.~Knapp\altaffilmark{\ref{Princeton}},
Robert H. Lupton\altaffilmark{\ref{Princeton}},
Jeffrey A. Munn\altaffilmark{\ref{USNO}},
Donald P.~Schneider\altaffilmark{\ref{PennState}}, 
Max Tegmark\altaffilmark{\ref{UPenn}}, and
Idit Zehavi\altaffilmark{\ref{UArizona}} 
}

\setcounter{address}{1}
\altaffiltext{\theaddress}{
\stepcounter{address}
Center for Cosmology and Particle Physics, Department of Physics, New
York University,  4 Washington Place, New York, NY 10003
\label{NYU}}
\altaffiltext{\theaddress}{
\stepcounter{address}
Princeton University Observatory, Princeton,
NJ 08544
\label{Princeton}}
\altaffiltext{\theaddress}{
\stepcounter{address}
Apache Point Observatory,
2001 Apache Point Road,
P.O. Box 59, Sunspot, NM 88349-0059
\label{APO}}
\altaffiltext{\theaddress}{
\stepcounter{address}
Institute for Cosmic Ray Research, University of
Tokyo, Midori, Tanashi, Tokyo 188-8502, Japan
\label{CosmicRay}}
\altaffiltext{\theaddress}{
\stepcounter{address}
Department of Astronomy, University of Washington,
Box 351580,
Seattle, WA 98195 
\label{UW}}
\altaffiltext{\theaddress}{
\stepcounter{address}
U.S. Naval Observatory,
3450 Massachusetts Ave., NW,
Washington, DC  20392-5420
\label{USNO}}
\altaffiltext{\theaddress}{
\stepcounter{address}
Department of Physics, University of Pennsylvania, Philadelphia,
PA 19104
\label{UPenn}}
\altaffiltext{\theaddress}{
\stepcounter{address}
Department of Astronomy and Astrophysics,
The Pennsylvania State University,
University Park, PA 16802
\label{PennState}}
\altaffiltext{\theaddress}{
\stepcounter{address}
Department of Astronomy, University of Arizona, Tucson, AZ 85721
\label{UArizona}}

\begin{abstract}
Here we present the New York University Value-Added Galaxy Catalog
(NYU-VAGC), a catalog of local galaxies (mostly below $z
\approx 0.3$) based on a set of publicly-released surveys
matched to the Sloan Digital Sky Survey (SDSS) Data Release 2. The
photometric catalog consists of 693,319 galaxies, QSOs and stars;
343,568 of these have redshift determinations, mostly from the
SDSS. Excluding areas masked by bright stars, the photometric sample
covers 3514 square degrees and the spectroscopic sample covers 2627
square degrees (with about 85\% completeness). Earlier, proprietary
versions of this catalog have formed the basis of many SDSS
investigations of the power spectrum, correlation function, and
luminosity function of galaxies. Future releases will follow future
public releases of the SDSS. The catalog includes matches to the
Two-Micron All Sky Survey Point Source and Extended Source catalogs,
the IRAS Point Source Catalog Redshift Survey, the Two-degree Field
Galaxy Redshift Survey, the Third Reference Catalog of Galaxies, and
the Faint Images of the Radio Sky at Twenty-centimeters survey.  We
calculate and compile derived quantities from the images and spectra
of the galaxies in the catalogs (for example, $K$-corrections and
structural parameters for galaxies). The SDSS catalog presented here
is photometrically calibrated in a more consistent way than that
distributed by the SDSS Data Release 2 Archive Servers and is thus
more appropriate for large-scale structure statistics, reducing
systematic calibration errors across the sky from $\sim 2\%$ to about
$\sim 1\%$. We include an explicit description of the geometry of the
catalog, including all imaging and targeting information as a function
of sky position. Finally, we have performed eyeball quality checks on
a large number of objects in the catalog in order to flag errors (such
as errors in deblending). This catalog is complementary to the SDSS
Archive Servers, in that NYU-VAGC's calibration, geometrical
description, and conveniently small size are specifically designed for
studying galaxy properties and large-scale structure statistics using
the SDSS spectroscopic catalog.
\end{abstract}


\section{Motivation}
\label{motivation}

New, large galaxy datasets such as the Sloan Digital Sky Survey (SDSS;
\citealt{york00a}) and the Two-Micron All Sky Survey (2MASS;
\citealt{skrutskie97a}) give 
astronomy a view of the local universe with an unprecedented
combination of completeness and detail. These surveys promise to
refine our understanding of properties of galaxies, their
relationships with environment, and their evolution.

However, these catalogs consist of terabytes of data and are therefore
currently too unwieldy simply to download onto a small workstation and
investigate directly. Furthermore, simple matched catalogs between
many surveys exist only in the form of web interfaces such as
NED\footnote{{\tt http://nedwww.ipac.caltech.edu/}}.  As indispensable
as such interfaces are for studying a small number of individual
objects on which one desires all the knowledge in the literature, they
are not ideal for huge batch jobs designed to run automatically on
relatively homogeneous data sets. Finally, none of the databases
interfacing to these huge catalogs yield any expression of the window
function on the sky of the included surveys, leaving this critical
determination to the user.

For these reasons, we have created a galaxy redshift catalog designed
to aid in the study of the local universe. Early versions of this
catalog have proven invaluable to the SDSS team, as they form the
basis for the work studying the luminosity function, power spectrum,
correlation function, and a number of other galaxy property and galaxy
clustering statistics; {\it e.g.} \citet{blanton03c, tegmark04a,
zehavi02a, hogg03b, hoyle03a, shen03a, baldry04a, pope04a}.  Non-SDSS
investigators have also used public data extracted from this catalog
to investigate galaxy properties and their evolution; {\it e.g.}
\citet{trujillo04a, rudnick03a}.  The catalog is small enough (tens of
Gbytes) to easily store on any modern machine, it contains matches
between many major public catalogs, and it contains an explicit
description of its window function. The current version of the catalog
uses the SDSS imaging survey as its basis, matching sources in other
catalogs to that master list, and only tracking the SDSS geometrical
information. In the future we plan on incorporating other surveys more
fully by explicitly including their geometrical information.

For most of the catalogs in the sample, we simply use the official
public releases. However, we give the SDSS (of which the authors are
participants) a special status in this catalog. First, it is the only
survey for which we track detailed information about the window
function.  Second, for the SDSS we use an independent set of
reductions of the public data (Schlegel et al. in preparation).  These
reductions are significantly improved over the reductions made
available through the Data Archive Server (DAS) or Catalog Archive
Server (CAS) on the SDSS Data Release 2 (DR2; \citealt{abazajian04a})
web site\footnote{{\tt http://sdss.org/dr2/}}.  The essential
improvement is that errors in the large-scale relative photometric
calibration of the survey have a lower variance and less structure,
making the catalog more appropriate for studies of large-scale
structure (as we demonstrate below). In addition, these reductions
include an explicit expression for the survey geometry which the
Archive Servers do not. Secondarily, the naming and unit conventions
in our catalog are different, as we describe in this paper. For these
reasons, expect the quantities for objects in this catalog and the
corresponding objects in the Archive Server catalogs to be very
similar but not identical.

This paper describes only the outline and general principles of the
catalog. We leave the detailed documentation to a set of regularly
updated web pages listed in Table \ref{webpages}.  The paper contains:
in Section
\ref{window}, a short description of the geometrical expressions used
in the catalog and how to use them; in Section \ref{constituents}, a
description of the constituent catalogs and how they are matched,
focusing on the SDSS; in Section
\ref{other}, a description of other derived quantities; in Section
\ref{lowz}, the presentation of a low redshift galaxy catalog based on
this data; in Section \ref{tools}, a short desription of some of the
tools we have used to create this dataset which might be useful to the
user as well; and in Section \ref{summary}, a summary.

Note that throughout we use the environmental variable \vagcdir{} to
denote the root location of the NYU-VAGC; see the online documentation
listed in Table \ref{webpages} for the actual root
location. Similarly, for the large-scale structure samples we describe
below, we use the environmental variable \lssdir{}. 

\section{Survey window geometries}
\label{window}

In order to use galaxy surveys in a statistically meaningful way, we
must describe their geometry on the sky. Doing so is important in
order to determine the size of the volume surveyed as well as to
determine in what area of sky a particular type of object {\it could
have been} observed. However, astronomers have no standard way of
expressing such geometrical information. Here we describe briefly the
extremely general and compact system we use, introduced by
\citet{hamilton03a}. Currently, we have described the SDSS geometry 
using this method, but have not included geometrical information for
the other catalogs in the NYU-VAGC.

Following \citet{hamilton03a}, we store the geometry on the sky as a
set of disjoint convex spherical polygons.  Spherical polygons have
several advantages: they are easy to express, it is easy to determine
whether a point is inside or outside, and there exist relatively
simple methods to transform them into spherical harmonic components
(\citealt{tegmark02a}).  Furthermore, they can express a variety of
shapes on the sky. Since the window functions of surveys, in
particular of the SDSS survey as described below, are generally
complex combinations of (for example) rectangles corresponding to
imaging coverage and (in the case of the SDSS) circles corresponding
to the spectroscopic coverage, a flexible expression of the geometry
is extremely useful.  The polygons are defined as the intersection of
a set of ``caps'' in the manner of Andrew Hamilton's product {\tt
mangle} (see Table \ref{webpages}). Each cap is defined as a part of a
spherical surface separated out by slicing of the sphere by a
plane. Such a slicing cuts the sphere into two parts, and therefore
choosing a cap means picking one of these parts.

In practice, we can fully specify a cap using four numbers (three
independent ones): the unit vector $\vv{\hat x}$ perpendicular to the
plane slicing the sphere and the quantity $c_m = \pm (1 -\cos
\theta)$, where $\theta$ is the polar angle defining the angular cap
radius. A positive $c_m$ indicates that a direction $\vv{a}$ is inside
the cap if
\begin{equation}
1 - \vv{x}\cdot\vv{a} < |c_m|.
\end{equation}
A negative $c_m$ indicates that a point is inside the cap if 
\begin{equation}
1 - \vv{x}\cdot\vv{a} > |c_m|.
\end{equation}

For the value-added catalog we define the Cartesian coordinates $x_i$
such that:
\begin{eqnarray}
\label{cartesian}
x_0 &=& \cos \delta \cos \alpha, \cr
x_1 &=& \cos \delta \sin \alpha, \mathrm{~and}\cr
x_2 &=& \sin \delta, 
\end{eqnarray}
where $\alpha$ and $\delta$ refer to the J2000 right ascension and
declination.

The polygons are described in FITS files of the form:
\begin{quote}
{\tt survey\_geometry.fits} 
\end{quote}
where ``survey'' indicates which survey we are describing. 
We also usually include ASCII versions in the standard polygon format
of \citet{hamilton03a} in an {\tt ascii} subdirectory
wherever the FITS version is found:
\begin{quote}
{\tt ascii/survey\_geometry.ply}\\
{\tt ascii/survey\_geometry\_info.dat}
\end{quote}
where the ``info'' file contains auxiliary columns describing
properties associated with each area of sky; in practice what these
properties are varies depending on the survey in question.  These
files contain the geometrical description of each polygon.  The format
is described on the {\tt mangle} homepage (see Table \ref{webpages}).
The description of how these structures are stored in the FITS files
is on the {\tt vagc} web page (see Table \ref{webpages}) and tools
exist in {\tt idlutils} to read them into IDL structures.


It is often necessary to combine two sets of polygons describing two
different surveys: for example, the SDSS imaging survey and the SDSS
spectroscopic survey. In order to do so, we use a procedure known as
``balkanization'' (\citealt{hamilton03a}). The concept is simple,
though the implementation is not. Imagine plotting both sets of
polygons on the same page. The lines plotted now define a new set of
disjoint polygons bounded by caps, which we refer to as balkans. Each
balkan is either entirely inside both surveys or only inside one
survey. We can express the intersection of the two catalogs as the set
of balkans entirely inside both surveys. The code presented by
\citet{hamilton03a} (see the web site listed in Table \ref{webpages}) 
is capable of finding this set of balkans.

\section{Constituent catalogs}
\label{constituents}

Here we describe the set of catalogs included in the NYU-VAGC: the
SDSS, FIRST, 2MASS, 2dFGRS, IRAS PSC$z$, and the RC3.

\subsection{SDSS}

The primary catalog is the SDSS.  The SDSS is acquiring $ugriz$ CCD
imaging of $10^4~\mathrm{deg^2}$ of the Northern Galactic sky, and,
from that imaging, selecting $10^6$ targets for spectroscopy, most of
them galaxies with $r<17.77~\mathrm{mag}$
\citep[e.g.,][]{gunn98a,york00a,abazajian03a}.  Automated software
performs all of the data processing: astrometry
\citep{pier03a}; source identification, deblending and photometry
\citep{lupton01a}; photometricity determination \citep{hogg01a};
calibration \citep{fukugita96a,smith02a}; spectroscopic target
selection \citep{eisenstein01a,strauss02a,richards02a}; spectroscopic
fiber placement \citep{blanton03a}; and spectroscopic data reduction.
An automated pipeline measures the redshifts and classifies the
reduced spectra (\citealt{stoughton02a}; Schlegel et al., in
preparation).

\begin{figure}
\figurenum{\fignum}
\plotone{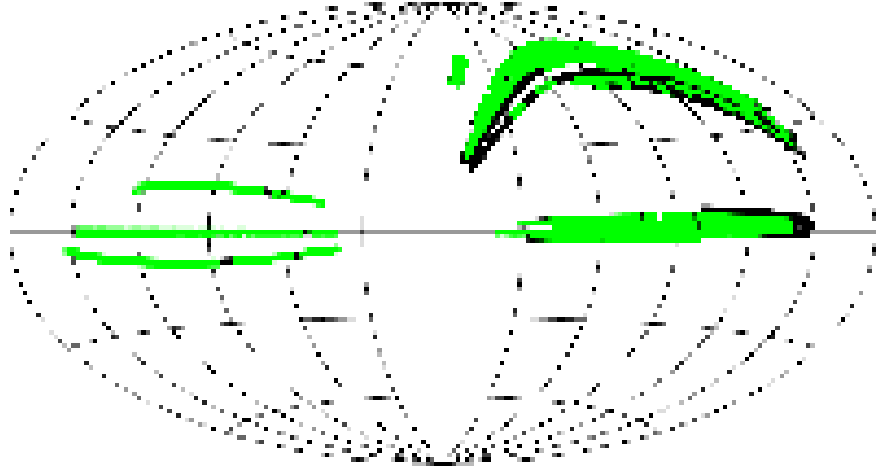}
\caption{\label{sdssim} Distribution on the sky of the NYU-VAGC in
  Equatorial coordinates. The center of this plot is the direction
  $\alpha=270$, $\delta=0$. The black indicates the coverage of the
  imaging, the green indicates the location of the spectra. The survey
  contains 3514 sq deg of imaging and 2627 sq deg of spectroscopy.}
\end{figure}

The NYU-VAGC catalog corresponds to public data up to the SDSS DR2.
We include three separate catalogs from SDSS DR2: the SDSS imaging
catalog, the SDSS tiling catalog (a description of the targets in the
imaging catalog for which we have a well-defined completeness), and
the SDSS spectroscopic catalog.  The top panel of Figure \ref{sdssim}
shows the geometry of the portion of the SDSS imaging survey (DR2)
released here, which includes 3514 sq deg of imaging.  The green
points in Figure \ref{sdssim} show the distribution on the sky of SDSS
spectra.

This list of objects includes stars, QSOs, and galaxies
together. Since the typical user of this catalog will be interested in
galaxies, we recommend using the {\tt VAGC\_SELECT} bitmask described
below if they want to select Main sample type galaxies with $m_r <
18$. Note that the criteria on which these are selected are slightly
broader than those used to target galaxies in the SDSS spectroscopy,
so some of these galaxies will not have spectra for that reason. In
addition, some of these ``galaxies'' will in fact be misidentified
stars or other sources (since our criteria will necessarily be less
reliable than that used by the SDSS target selection process). For
maximum reliability, the user can always trim our catalog to obey the
same criteria as the SDSS target selection described in
\citet{strauss02a}. 

\subsubsection{SDSS imaging catalog} 
\label{sdss_imaging}

The SDSS imaging catalog presented here includes photometric
reductions using {\tt PHOTO} {v5\_4} (Lupton et al. in preparation),
and is described in \citet{abazajian04a}. However, unlike the data set
\citet{abazajian04a} presents, it is calibrated using overlaps of
SDSS runs (Schlegel et al, in preparation). This procedure results in
a more consistent large-scale calibration of the survey. In addition,
as we describe below, the primary area of the survey is defined
differently (and thus includes a somewhat larger fraction of the total
area of the DR2 imaging). 

We use some SDSS jargon below to discuss the organization of the data.
As described in the \citet{stoughton02a}, a ``run'' is a sequence
number assigned to a particular drift scan observation. A ``camcol''
(between 1 and 6) indicates a set of $ugriz$ CCDs in the focal
plane. In each run and camcol, the imaging data are divided in the
scan direction into $\sim 10$ arcmin ``fields'' for convenience.

Only a small number of the objects in the full SDSS catalog are
included here. Specifically, we include only:
\begin{enumerate}
\item A sample of objects similar to the SDSS Main galaxy sample
described by \citet{strauss02a} selected from the most recent (v5\_4)
version of the imaging reductions. We have changed some parameters to
be slightly more inclusive:
\begin{enumerate}
\item We extend the extinction-corrected Petrosian magnitude limit 
from $r=17.77$ to $r=18$, in order to include spectroscopic objects which would
otherwise scatter outside the original flux limits due to changes in
calibration since targeting.
\item We extend the star-magnitude separation from 0.3 mag to 0.2 mag (expressed as
$m_{\mathrm{PSF}}-m_{\mathrm{model}}$ in the $r$-band; see
\citealt{strauss02a} for details). This change includes a small number
of galaxies not included in the targeting, but also introduces a
number of stellar sources into the catalog. 
\item We extend the bright fiber magnitude limit from $g,r,i > 15, 15,
14.5$ to $g,r,i > 12$
(effectively turning it off). 
\item  We turn off the exclusion of small, bright objects (with $R_{50}
< 2''$ and $m_P < 15$). This change results in the inclusion of a small
number of binary stars. 
\end{enumerate}
\item The closest object within $2''$ of the position of each SDSS
spectrum. In this match, we in fact give priority to objects that pass
the previous criterion to be a spectroscopic galaxy target in the
latest reductions. One might worry that this would introduce some sort
of flux bias in the sample, since these targets are bright relative to
the typical object, but in fact any pairs of objects with separations
less than $2''$ are likely to be spurious in any case. Thus, taking
the brightest one is usually the correct choice.
\item The closest object within $2''$ of the position of each Main
sample, QSO, or Luminous Red Galaxy spectroscopic target from the
target version of the reductions which was fed to the tiling software
(a ``tiled target''), whether or not they had spectra taken. These
targets were selected based on earlier reductions of the imaging which
differ in significant ways from the latest reductions.
\end{enumerate}
We have retained the Petrosian half-light $r$-band surface brightness
limit at $\mu_{50} = 24.5$. Note that the new objects included by the
extensions described above are unlikely to have {\it spectra}.

In fact, because of catalog errors, there remain a handful of pairs of
objects within $2''$ of each other (12 out of 693,331). After removing
these duplicates, we include one observation of each object in the
file:
\begin{quote}
\vagcdir{\tt /object\_sdss\_imaging.fits}
\end{quote}
which contains a trimmed set of the photometric measurements (the
exact list is described on the web page listed in Table
\ref{webpages}). 
We use the environmental variable \vagcdir{} to
denote the root location of the NYU-VAGC; see the online documentation
for the actual root location. 
In addition, we store all of the
photometric information for this set of objects in another set of
files of the form:
\begin{quote}
\vagcdir{\tt /sdss/parameters/calibObj-[run]-[camcol].fits},
\end{quote}
where, as indicated, objects in separate ``runs'' and ``camcols'' are
kept in separate files. 
The {\tt object\_sdss\_imaging} file
contains for each object its run and camcol, as well as its
zero-indexed position in the corresponding {\tt calibObj} file,
allowing quick access to the full photometric information.

The geometry of the SDSS imaging survey is in a file:
\begin{quote}
\vagcdir{\tt /sdss/sdss\_imaging\_geometry.fits}
\end{quote}
stored in the form of spherical polygons. We create a list of polygons
that describe the primary area of each imaging field. For each
polygon, we give the SDSS imaging field (that is, its {\tt run}, {\tt
camcol}, and {\tt field}) which we consider ``primary'' for the
purpose of resolving duplicate observations.

We should note here some points about the quantities in the SDSS
data we present (some of which differ from data distributed by the
Archive Servers):
\begin{enumerate}
\item All fluxes are given in ``nanomaggies'' $f$,
which represent the flux (multiplied by $10^9$, as the prefix ``nano''
implies) relative to that of the AB standard source with $f_\nu =
3631$ Jy (\citealt{oke68a}). They are related to standard astronomical
magnitudes $m$ by the formula:
\begin{equation}
\label{nmgy}
m = 22.5 - 2.5 \log_{10} f 
\end{equation}
\item The zeropoints of the system are the same as that for the data
on the Archive Servers, so that the conversion to AB magnitudes is the
same as that given by \citet{abazajian04a}.  
\item Uncertainties are expressed in terms of the inverse variance
(usually used for calibrated quantities, with the suffix {\tt \_IVAR})
or in terms of the standard deviation (usually used with uncalibrated
quantities, with the suffix {\tt ERR}). 
\item The column {\tt VAGC\_SELECT} is a bitmask that yields 
information on how each imaging object was selected, with the following bits:
\begin{itemize}
\item[0:] near tiled target
\item[1:] near spectrum
\item[2:] pass the Main sample galaxy criteria (with the adjustments listed
above)
\end{itemize}
Thus, an object which passed the galaxy criteria, and was near a tiled
target, and was near a spectrum, would have bits 0, 1, and 2 all set,
resulting in a numerical value of 7 for {\tt VAGC\_SELECT}. 
\item In addition to the local sky determination (the $100''$ by $100''$
median smoothed sky in {\tt skyflux}) we provide the sky estimate for
the current $9.8'$ by $13.5'$ field as a whole (``global'' sky) in the
parameter {\tt psp\_skyflux} (in nanomaggies per arcsec$^2$).
\item A crude bulge-to-disk decomposition exists for each object
processed by {\tt PHOTO}, which simply consists of taking the best fit
de Vaucouleurs model, the best fit exponential model, linearly
combining them and refitting for the amplitudes of the models (see
\citealt{abazajian04a} for details). The
fraction of the flux assigned to de Vaucouleurs model (the ``bulge
fraction,'' if you will allow it) is put in the column called {\tt
fracpsf}.
\end{enumerate}

In addition to these changes of form, there is a fundamental
difference between the Princeton reductions used here and the data
available on the SDSS Archive Servers --- the relative photometric
calibration (Schlegel et al. in preparation). The Archive Server
reductions use the calibration procedure described in
\citet{abazajian03a}, which involves comparing counts measured on the
2.5m telescope, to those measured on a nearby 0.5m photometric
telescope on a slightly different filter system, to the fluxes of a
set of primary standard stars (on a yet different filter
system). Instead of this procedure, Schlegel et al. (in preparation)
take advantage of the wide angular baseline provided by the drift
scanning and of the large number of overlapping observations. This
combination results in many exposures of the same stars taken on
different nights. One can then use these multiple observations to fit
for the calibration parameters (system response, airmass term, flat
fields) as a function of time by minimizing the differences between
the resulting inferred fluxes of the multiply-observed stars. This
procedure, denoted ``ubercalibration,'' takes advantage of the fact
that the system is photometrically stable within each drift scan run,
and uses that to tie all the runs together using their overlapping
observations.

\begin{figure}
\epsscale{0.7}
\figurenum{\fignum}
\plotone{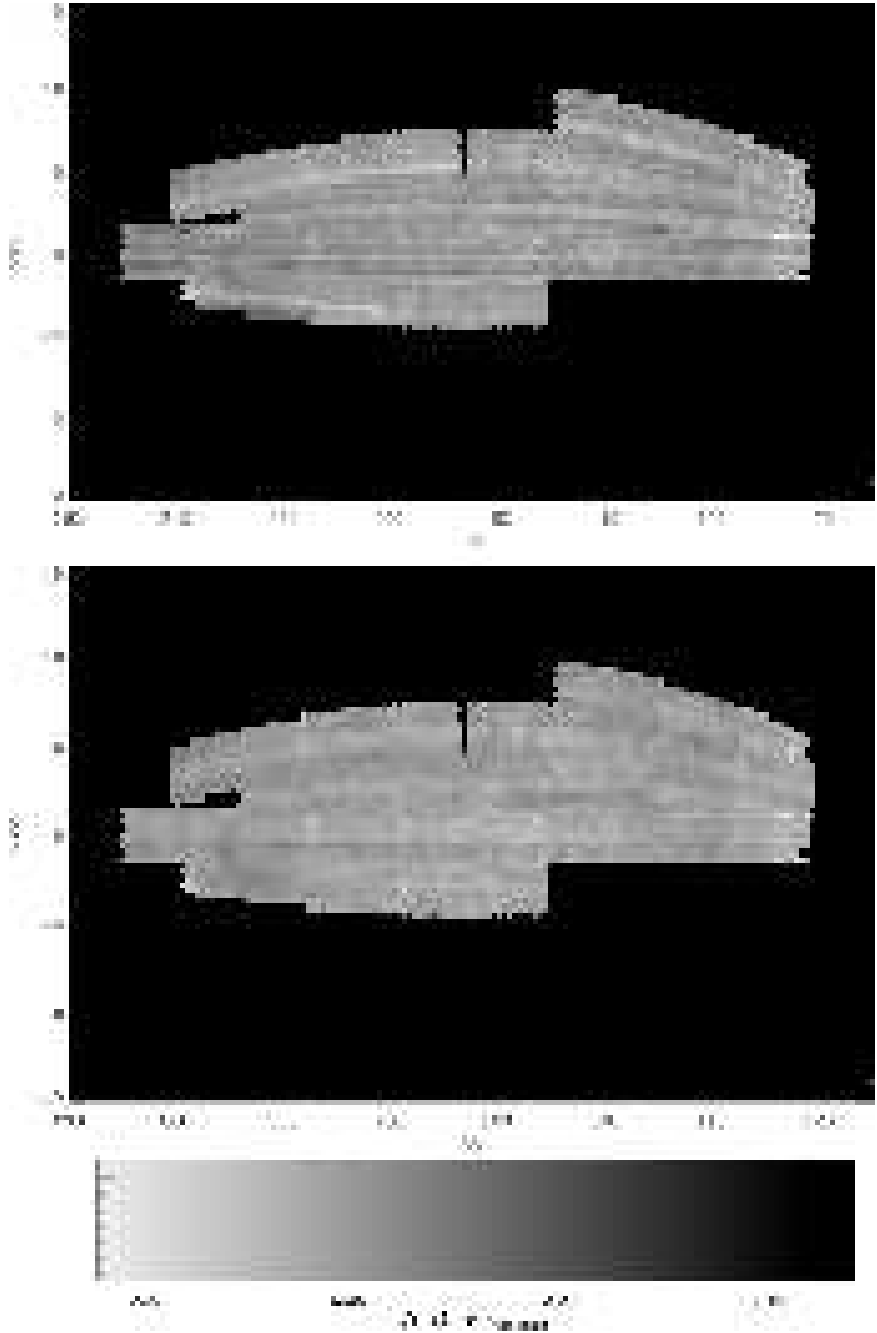}
\caption{\label{bluetip} The top two panels show the $r-i$ color of
the bluest stars in the magnitude range $16 < m_r < 18.5$ in each
contiguous set of twenty fields in each run of the SDSS (all
magnitudes extinction-corrected according to the dust maps of
\citealt{schlegel98a}). The bottom panel shows the calibration of the
scale. Only the Northern Equatorial data is shown.  The top panel
shows this quantity for the data calibrated to the SDSS standard
system using the photometric telescope. The middle panel shows the
same for the ubercalibrated data, as described in the text. }
\end{figure}

As a demonstration that the procedure results in a lower variance in
large-scale errors in the calibration, consider Figure \ref{bluetip}.
The greyscale in the top two panels shows the $r-i$ color of the
bluest stars in the magnitude range $16 < m_r < 18.5$ in each
contiguous set of twenty fields in each run of the SDSS (all
magnitudes extinction-corrected according to the dust maps of
\citealt{schlegel98a}). We only show one section of the SDSS coverage
on the Northern Equator, but similar results hold elsewhere. This
blue-tip color varies smoothly across the sky due to metallicity
gradients in the Galactic stellar halo, but has little small scale
structure. The top panel shows this quantity for data we have
calibrated to the SDSS standard system using the photometric telescope
(though these results are {\it not} identical to those in the SDSS
Archive). The bottom panel shows the same for the ubercalibrated
data. Both panels reveal the dependence of stellar metallity on
Galactic latitude, as well as some large-scale features in the stellar
distribution. Clearly, the stripy variations in the top panel, which
are errors in the SDSS calibration, are greatly reduced in the bottom
panel (though not eliminated). The 5-sigma clipped standard deviation
in $r-i$ color over the whole SDSS area is reduced from 0.02 to 0.01
mag between the PT calibration and the ubercalibration (these numbers
{\it include} the variation of the stellar populations). The rms
variations of $r-i$ within several degree scale patches is about 0.007
mag. These results suggest that ubercalibration is a significant
improvement over the standard calibration {\it and} that the
calibration is good to about 1\%. The SDSS is taking long scans across
the entire survey in the Northern Galactic Cap, as well as scans that
connect the three separated stripes in the Southern Galactic Cap, that
will reduce the systematic errors even further.

Our web site (see Table \ref{webpages}) has full documentation of the
structure of the files described in this and subsequent sections and
of all of the parameters they contain.

\subsubsection{SDSS tiling catalog}
\label{tiling}

\begin{figure}
\figurenum{\fignum}
\plotone{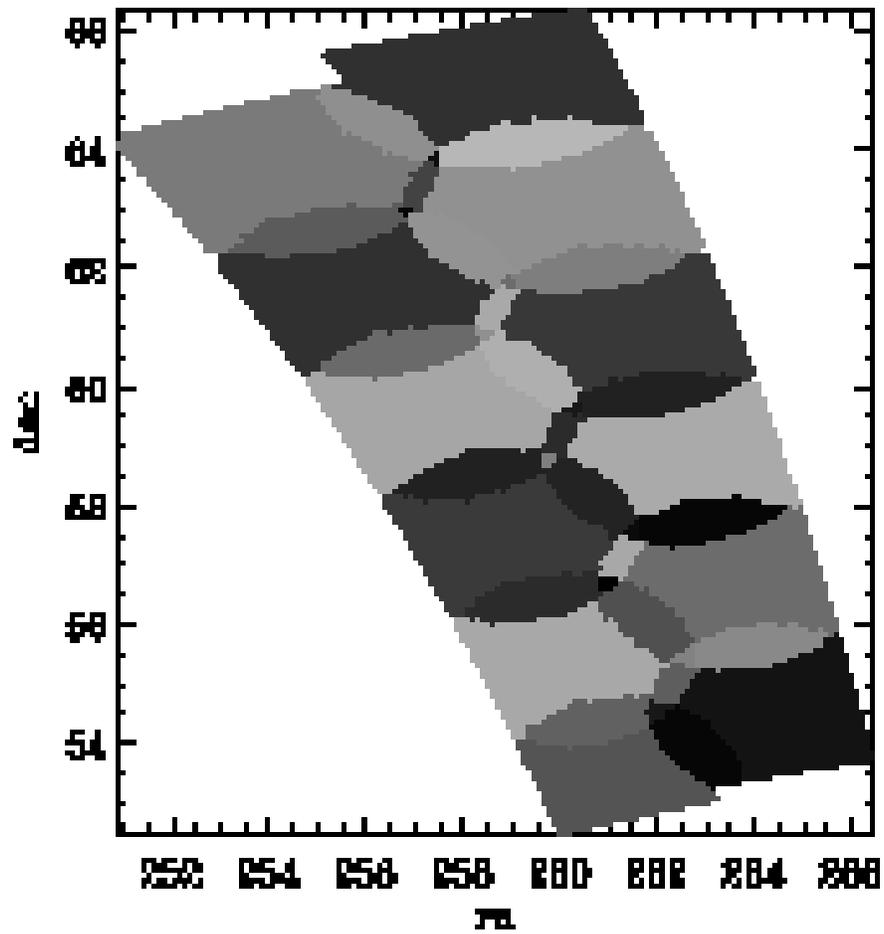}
\caption{\label{tilingregion} An example of a ``tiling region'' in the
survey (Tiling Region 7). Each shade of grey indicates the coverage of
a given ``sector'' (as defined in the text, a region covered by a
unique set of tiles).}
\end{figure}

The primary SDSS spectroscopic program proceeds in the following
manner. Based on a set of images for which we have selected targets
(using the algorithms in \citealt{strauss02a},
\citealt{eisenstein01a}, \citealt{richards02a}, and
\citealt{stoughton02a}), the SDSS defines
a ``tiling region.'' For example, for Tiling Region 7 in the SDSS,
Figure \ref{tilingregion} shows the region we defined.  Given the
tiling region and the set of targets within it we determine the
location of spectroscopic tiles of radius 1.49 deg and decide to which
targets to assign fibers (\citealt{blanton03a}). This procedure
defines a set of 1.49 deg radius circles on the sky corresponding to
the tiles. The intersection of the rectangles describing the tiling
region and the circles describing the tiles defines the geometry of
the tiling region. In this geometry we define ``sectors,'' each of
which consists of a set of spherical polygons which could have been
observed by a unique set of tiles. These sectors are the appropriate
regions on which to define the completeness of the survey. See the DR2
web site\footnote{{\tt http://www.sdss.org/dr2/products/tiling}} for
more complete documentation on tiling. In Figure \ref{tilingregion},
we have given each sector a different shade of grey.

The union of the tiling regions defines the geometry of the
spectroscopic survey as a whole.  Note that this geometry is not,
generally, as simple as the total area covered by the tiles. For
example, some regions are within 1.49 deg of the center of a tile, but
the tile was created before spectroscopic targets for that region had
been selected. This fact of life results in gaps in the survey which
we track in our geometrical description of the survey.

The set of polygons describing the tiling geometry is in:
\begin{quote}
\vagcdir{\tt /sdss/sdss\_tiling\_geometry.fits}
\end{quote}
This file yields the sector to which each polygon belongs. The
properties of the sectors are given in:
\begin{quote}
\vagcdir{\tt /sdss/sdss\_sectorList.par},
\end{quote}
which yield which tile each sector belongs to. Finally, the centers of
each tile are given in 
\begin{quote}
\vagcdir{\tt /sdss/tileFull.par}.
\end{quote}

We match the set of objects which we input into the tiling program to
the nearest imaging object within $2''$ in the {\tt
object\_sdss\_imaging} catalog and put the results in the file:
\begin{quote}
\vagcdir{\tt /object\_sdss\_tiling.fits}
\end{quote}
Each entry in this file refers to the corresponding entry in the {\tt
object\_sdss\_imaging} file; e.g. row number 3 in one file refers to
the same object as row number 3 in the other. The
full set of tiled objects (including those that do not match any of
the imaging objects) is included in the file:
\begin{quote}
\vagcdir{\tt /sdss/sdss\_tiling\_catalog.fits}
\end{quote}
The {\tt object\_sdss\_tiling.fits} file has the column {\tt
sdss\_tiling\_tag\_primary} which gives the zero-indexed row number of
the object in the {\tt sdss\_tiling\_catalog.fits} file. Unmatched
objects have {\tt sdss\_tiling\_tag\_primary == -1} (and the rest of
the columns for such rows are set to appropriate null values). 

The reader may wonder why there would be any unmatched objects. The
reason is the photometric reduction code has changed over time, in
particular the deblending algorithm has changed. For this reason,
there are occasionally objects found in an old reduction which have no
corresponding object within $2''$ in the new reductions, because the
set of detected pixels in that region has been deblended differently
by the two versions of the code.

Again, the structure and contents of these files are described on the
web site.

\subsubsection{SDSS spectroscopic catalog}

For this catalog we use the reductions of the SDSS spectroscopic data
performed by Schlegel et al. (in prep) using their reduction code {\tt
idlspec2d}, which extracts the spectra and finds the redshifts. The
redshifts found by {\tt idlspec2d} are almost always (over 99\% of the
time for Main galaxy sample targets) identical to the redshifts found
by an alternative pipeline used for the SDSS Archive Servers (SubbaRao
et al. in prep).

We match the set of objects for which we have SDSS spectra to the
nearest imaging object within $2''$ in the {\tt object\_sdss\_imaging}
catalog and put the results in the file:
\begin{quote}
\vagcdir{\tt /object\_sdss\_spectro.fits}
\end{quote}
Each entry in this file refers to the corresponding entry in the {\tt
object\_sdss\_imaging} file; e.g. row number 3 in one file refers to
the same object as row number 3 in the other. 
The full set of spectra (including those that do not match any of the
imaging objects) is included in the file:
\begin{quote}
\vagcdir{\tt /sdss/sdss\_spectro\_catalog.fits}
\end{quote}
The {\tt object\_sdss\_spectro.fits} file has the column {\tt
sdss\_spectro\_tag\_primary} which gives the zero-indexed row number
of the object in the {\tt sdss\_spectro\_catalog.fits} file. Unmatched
objects have \\ {\tt sdss\_spectro\_tag\_primary == -1} (and the rest
of the columns for such rows are set to appropriate null values).
Note that in addition to the issues regarding deblending in different
reductions mentioned in the previous subsection, a number of the
spectra are sky spectra which are placed randomly on the sky and will
never correspond across reductions.

We do not provide any geometrical description of the catalog beyond
the locations of each fiber (each one is $3''$ diameter). 

The quantities in these files are documented at the spectroscopic
reduction web page (see Table \ref{webpages}).

\subsection{FIRST}

Using the Very Large Array, the Faint Images of the Radio Sky at
Twenty-centimeters (FIRST; \citealt{becker95a}) survey has mapped
10,000 square degrees of the Northern Sky overlapping the SDSS with a
detection limit about 1 mJy and a resolution of 5$''$.  For each object
in {\tt object\_sdss\_imaging} we find the matching object in the
FIRST catalogs within 3 arcsec. The columns with the prefix {\tt
FIRST} in the files:
\begin{quote}
\vagcdir{\tt /sdss/parameters/calibObj-\$run-\$camcol.fits}
\end{quote}
contain the FIRST results. The columns are described in detail in the
Princeton photometric reduction web site listed in Table
\ref{webpages}. 

\subsection{2MASS}
\label{twomass}

2MASS (\citealt{cutri00a}) is an all-sky map in $J$, $H$, and
$K_s$. Two catalogs have been developed from this map; the Point
Source Catalog (PSC, complete to roughly $K_s\sim 15$, Vega-relative)
and the Extended Source Catalog, that is, the galaxy catalog (XSC,
complete to roughly $K_s \sim 13.5$). For each object in {\tt
object\_sdss\_imaging} we find the matching object within 3
arcsec. The columns with the prefix {\tt TMASS} in the files:
\begin{quote}
\vagcdir{\tt /sdss/parameters/calibObj-\$run-\$camcol.fits}
\end{quote}
contain the 2MASS PSC data.  The columns are described in detail in
the Princeton photometric reduction web site listed in Table
\ref{webpages}.

In addition, we match the 2MASS Extended Source Catalog (described in
the Explanatory Supplement to the 2MASS All Sky Data Release; see
Table \ref{webpages}; \citealt{cutri00a}) to objects within $3''$ in
{\tt object\_sdss\_imaging} and put the results in the file:
\begin{quote}
\vagcdir{\tt /object\_twomass.fits}
\end{quote}
Each entry in this file refers to the corresponding entry in the {\tt
object\_sdss\_imaging} file; e.g. row number 3 in one file refers to
the same object as row number 3 in the other. The full set of 2MASS
XSC
objects (including those that do not match any of the imaging objects)
is included in the files:
\begin{quote}
\vagcdir{\tt /twomass/twomass\_catalog\_000.fits}\\
\vagcdir{\tt /twomass/twomass\_catalog\_001.fits}\\
\vagcdir{\tt /twomass/twomass\_catalog\_002.fits}\\
\vagcdir{\tt /twomass/twomass\_catalog\_003.fits}
\end{quote}
These files contain a subset of the columns listed by
\citet{cutri00a}. Most notably, below we will use the 
``extrapolated'' galaxy magnitudes from these files, as described by
\citet{jarrett03a}.  

We have {\it not} converted the numbers in these files from their
original Vega-relative meaning. Where necessary below, we will use the
conversions to AB:
\begin{eqnarray}
J_{\mathrm{AB}} = J_{\mathrm{Vega}}+0.91 \cr
H_{\mathrm{AB}} = H_{\mathrm{Vega}}+1.39 \cr
K_{s,\mathrm{AB}} = K_{s,\mathrm{Vega}}+1.85 
\end{eqnarray}
calculated by the {\tt kcorrect v3\_2} code presented by
\citet{blanton03b}, using the filter curves of \citet{cutri00a} and
the theoretical Vega flux presented by \citet{kurucz91a}.

\begin{figure}
\figurenum{\fignum}
\plotone{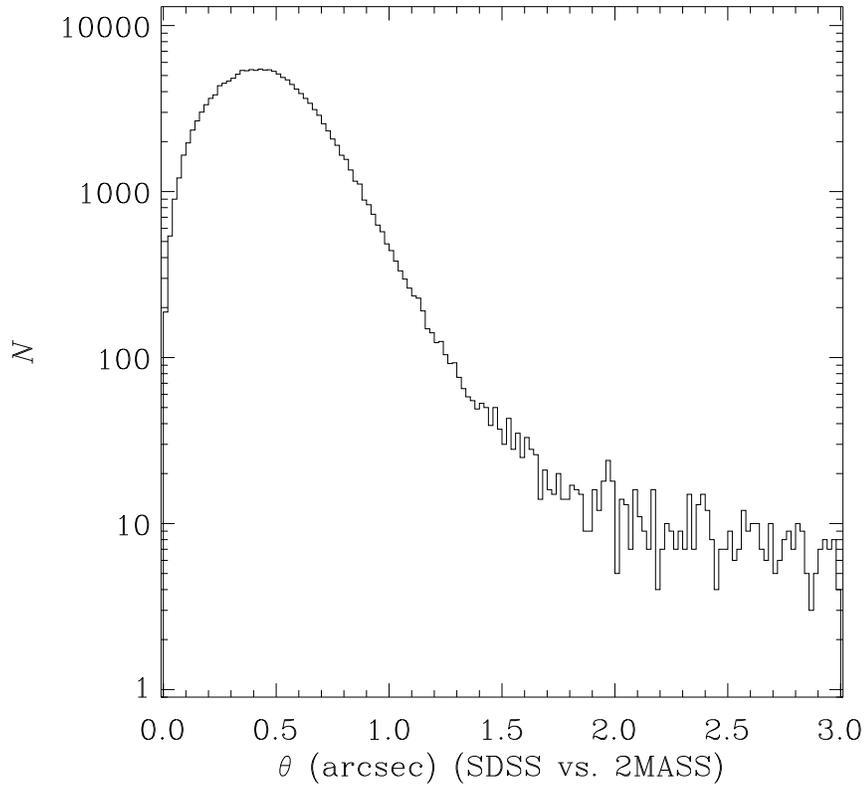}
\caption{\label{twomass_dist} Distribution of angular distances
between the matched objects in the SDSS and 2MASS (in the {\tt
object\_sdss\_imaging} and {\tt object\_twomass} catalogs). The
astrometry of the surveys agrees well.}
\end{figure}

Figure \ref{twomass_dist} shows the distribution of match distances
between the SDSS and 2MASS catalogs, showing the agreement in the
astrometry between these two catalogs (\citealt{pier03a,
finlator00a}).

\subsection{2dFGRS}

The 2dFGRS (\citealt{colless01a}) is a galaxy redshift survey using
the 2dF multi-object spectrograph, targeted off of the APM survey
(\citealt{maddox90a}). We match the 2dFGRS Final Data Release to
objects within $4''$ in {\tt object\_sdss\_imaging} and put the
results in the file:
\begin{quote}
\vagcdir{\tt /object\_twodf.fits}
\end{quote}
Each entry in this file refers to the corresponding entry in the {\tt
object\_sdss\_imaging} file; e.g. row number 3 in one file refers to
the same object as row number 3 in the other. The full set of 2dFGRS
objects (including those that do not match any of the imaging objects)
is included in the files:
\begin{quote}
\vagcdir{\tt /twodf/twodf\_catalog.fits}\\
\end{quote}

\begin{figure}
\figurenum{\fignum}
\plotone{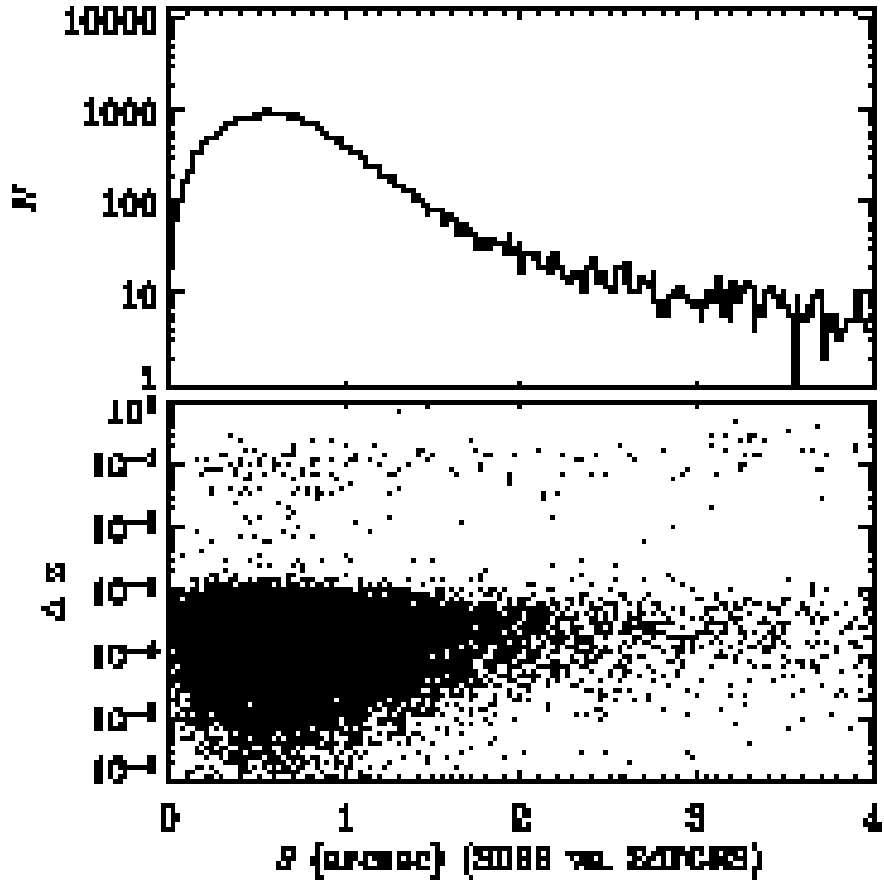}
\caption{\label{twodf_dist} {\tt Top panel:} Distribution of angular
distances between the matched objects in the SDSS and 2dFGRS (in the
{\tt object\_sdss\_imaging} and {\tt object\_twodf} catalogs). {\tt
Bottom panel:} Absolute redshift differences versus angular distances
for same matches. A small number (170) of these galaxies are large
outliers. In all such cases the SDSS pipeline either yields a clearly
correct redshift based on an eyeball examination of the spectra {\it
or} has flagged the spectrum as problematic with the {\tt ZWARNING}
flag. }
\end{figure}

The top panel of Figure \ref{twodf_dist} shows the distribution of
angular separations of our matches, and in the bottom panel (for
objects with redshifts in both catalogs) the (absolute) difference
between the SDSS and 2dFGRS redshifts versus the angular distance. We
limit our comparison to 2dFGRS redshifts with {\tt QUALITY} $\ge 3$
(the recommended criterion for a reliable redshift). There are around
27000 objects with redshifts in both catalogs; 94 of these are large
redshift discrepancies ($|\delta z| > 0.01$).  For about 83 of these
discrepancies, the SDSS redshift is clearly correct based on an
eyeball inspection of the extracted spectrum. For the remaining 11,
the SDSS redshift is flagged as poor by the spectroscopic reduction
software (using the {\tt ZWARNING} flag described on the spectroscopic
reduction web page listed in Table
\ref{webpages}).  

\subsection{IRAS PSC$z$}

The PSC$z$ is a redshift catalog of point sources in the IRAS survey
(\citealt{saunders00a}). Given the resolution of the IRAS survey, we
have matched each source to the nearest object within $40''$ in the
{\tt object\_sdss\_imaging} file, putting the results in the file:
\begin{quote}
\vagcdir{\tt /object\_pscz.fits}
\end{quote}
Each entry in this file refers to the corresponding entry in the {\tt
object\_sdss\_imaging} file; e.g. row number 3 in one file refers to
the same object as row number 3 in the other. The full set of PSC$z$
objects (including those that do not match any of the imaging objects)
is included in the files:
\begin{quote}
\vagcdir{\tt /pscz/pscz\_catalog.fits}\\
\end{quote}

\begin{figure}
\figurenum{\fignum}
\plotone{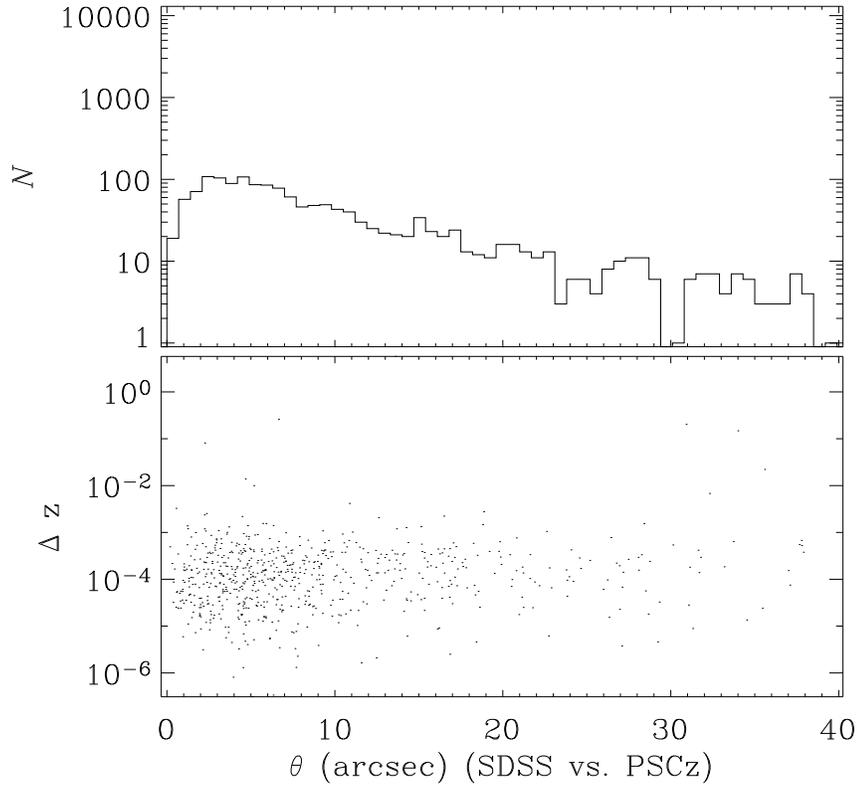}
\caption{\label{pscz_dist} Similar to Figure \ref{twodf_dist}, for the
SDSS and PSC$z$ (the {\tt object\_sdss\_imaging} and {\tt
object\_pscz} catalogs). The redshift differences are small even for
objects with large differences in astrometry, indicating that the
matches are reliable out to 40 arcsec.}
\end{figure}

Figure \ref{pscz_dist} shows in the top panel the distribution of
angular separations of our matches, and in the bottom panel (for
objects with redshifts in both catalogs) the (absolute) redshift
difference versus the angular distance. The redshift matches are good
in all cases.

\subsection{RC3}

The Third Reference Catalog of Galaxies (RC3) is a catalog of nearby
galaxies developed by
\citet{devaucouleurs91a}.  Because of the size of these galaxies (and
the fact that locations in the RC3 are occasionally only listed to the
nearest arcminute), we have matched each source to the nearest object
within $45''$ in the {\tt object\_sdss\_imaging} file, putting the
results in the file:
\begin{quote}
\vagcdir{\tt /object\_rc3.fits}
\end{quote}
Each entry in this file refers to the corresponding entry in the {\tt
object\_sdss\_imaging} file; e.g. row number 3 in one file refers to
the same object as row number 3 in the other. The full set of RC3
objects (including those that do not match any of the imaging objects)
is included in the files:
\begin{quote}
\vagcdir{\tt /rc3/rc3\_catalog.fits}\\
\end{quote}

\begin{figure}
\figurenum{\fignum}
\plotone{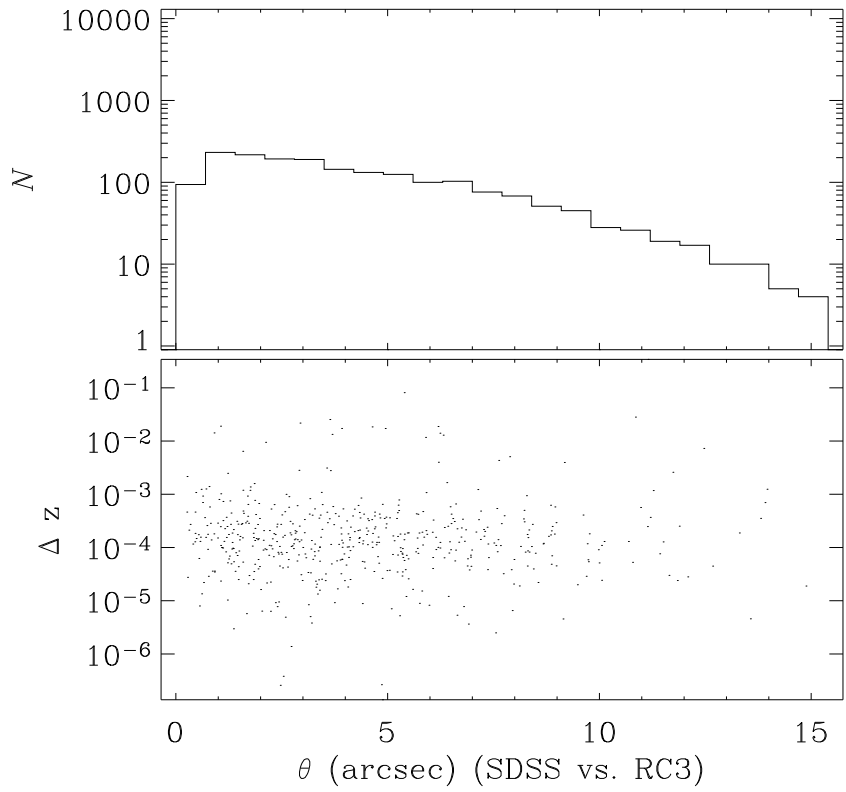}
\caption{\label{rc3_dist} Similar to Figure
\ref{twodf_dist}, 
for the SDSS and RC3 (the {\tt
object\_sdss\_imaging} and {\tt object\_rc3} catalogs). The matches
appear reliable out to 15 arcsec. }
\end{figure}

Figure \ref{rc3_dist} shows in the top panel the distribution of
angular separations of our matches, and in the bottom panel (for
objects with redshifts in both catalogs) the (absolute) redshift
difference versus the angular distance.  

\section{Additional quantities}
\label{other}

In addition to matches to external catalogs, we provide some extra
quantities measured from the NYU-VAGC catalog.

\subsection{Collision ``corrections''}
\label{collisions}

\begin{figure}
\figurenum{\fignum}
\plotone{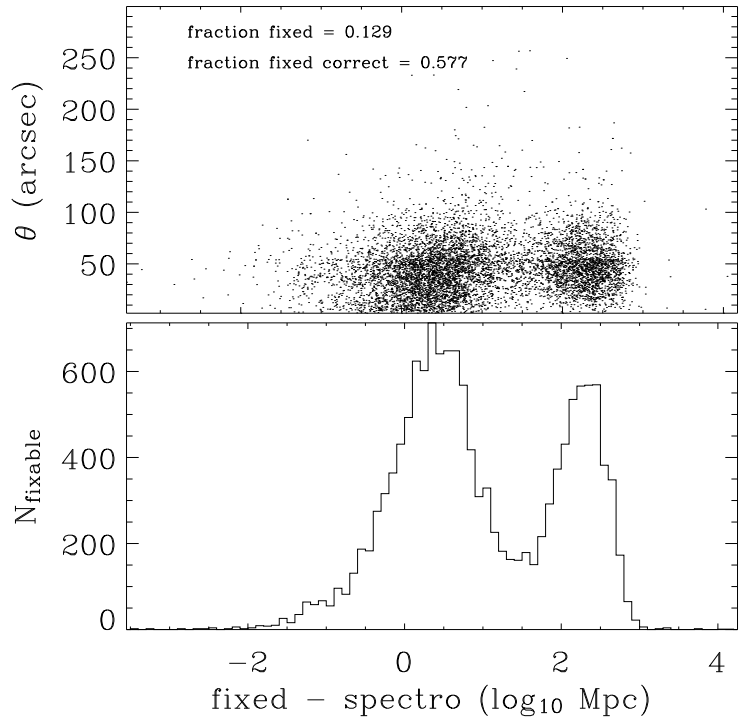}
\caption{\label{collisions.nearest} Top panel shows the distribution
of the redshift and angular differences of pairs of galaxies which
could have been corrected using the procedure of Section
\ref{collisions} but in fact have an available redshift. Bottom panel
shows the histogram of the distribution of redshift differences. }
\end{figure}

For the purposes of large-scale structure statistics with the SDSS, it
is necessary to account for the fact that some galaxies are missing in
the spectroscopic sample due to collided fiber constraints (no two
fibers on the same tile can be placed more closely than $55''$). We
can do so using the following procedure:
\begin{enumerate}
\item Group the galaxies according to a friends-of-friends procedure
with a $55''$ linking length.
\item For each galaxy which does not have a redshift in the SDSS data,
ask whether there is a galaxy (or galaxies) in its group with a
redshift.
\item If so, assign to the galaxy without a redshift that of the 
galaxy in the group which is closest on the sky and which has a
redshift.
\end{enumerate}
We put the results of this procedure into a file:
\begin{quote}
\vagcdir{\tt /collisions/collisions.nearest.fits}
\end{quote}
About 5--6\% of galaxies brighter than the flux limit need to be and
can be assigned a redshift using this criterion, as found previously
by \citet{zehavi02a}.  Judging from the cases which {\it could} have
been corrected but in fact had a redshift, about 60\% of the corrected
cases are within 10 $h^{-1}$ Mpc of the correct redshift. Figure
\ref{collisions.nearest} shows the distribution of redshift
separations and angular separations of such galaxies in the top
panel, and the histogram of redshift separations in the bottom
panel. 

We also have implemented a slight variant of this procedure, in which
we also require that the photometric redshift of the galaxy
(determined using {\tt kcorrect} {v3\_2}; \citealt{blanton03b}) which
was collided be within 0.05 of the spectroscopic redshift which we
want to assign to it. There is very little difference in the results;
we include them in the file
\begin{quote}
\vagcdir{\tt /collisions/collisions.photoz.fits}
\end{quote}
About 9\% of the objects are corrected, about 71\% of which are likely
to be within 10 $h^{-1}$ of the correct redshift (based on the cases
which could have been corrected but in fact had a redshift).

Finally, for completeness we include corresponding files without any
corrections at all:
\begin{quote}
\vagcdir{\tt /collisions/collisions.none.fits}
\end{quote}

\subsection{$K$-corrections}
\label{kcorrect}

We use the $K$-correction software {\tt kcorrect} {\tt v3\_2}
(\citealt{blanton03b}) to determine $K$-corrections for all of the
objects in the NYU-VAGC. We treat them all as if they are normal
galaxies; thus, the $K$-corrections are not going to be appropriate
for QSOs.

In the directory:
\begin{quote}
\vagcdir{\tt /kcorrect}
\end{quote}
we provide these estimates of the $ugrizJHK_s$ $K$-corrections and
absolute magnitudes of each object (using a $\Omega_0 =0.3$,
$\Omega_\Lambda = 0.7$ cosmology with $H_0 = 100$ $h$ km s$^{-1}$
Mpc$^{-1}$ for $h=1$). The $JHK_s$ fluxes all come from the 2MASS XSC
extrapolated magnitudes (\citealt{jarrett03a}) and have been converted
from the Vega system to the AB system as described in Section
\ref{twomass}. The files also contain the
Galactic-extinction corrected AB nanomaggies for each object. We
provide these for each set of collision corrections, for different
types of SDSS flux measurements, and for different rest-frame
bandpasses.

They are in files of the form:
\begin{quote}
\vagcdir{\tt /kcorrect/kcorrect.\$collision.\$flux.z\$bandshift.fits}
\end{quote}
where {\tt \$collision} refers to the type of collision correction
(that is {\tt none}, {\tt nearest}, or {\tt photoz}), {\tt \$flux}
refers to the type of flux used for the SDSS data (based on the prefix
used in the {\tt calibObj} files), and {\tt \$bandshift} refers to the
blueshift of the bandpasses we are correcting to.  As an example:
\begin{quote}
\vagcdir{\tt /kcorrect/kcorrect.none.petro.z0.10.fits}
\end{quote}
contains corrections for galaxies using no collision corrections,
Petrosian fluxes for the SDSS observations, and shifted to the
equivalent bandpass shapes at $z=0.10$.

\subsection{Sersic profile fits}

In the file
\begin{quote}
\vagcdir{\tt /sersic/sersic\_catalog.fits}
\end{quote}
we provide \Sersic\ fits to the azimuthally averaged radial profiles
of each object (\citealt{sersic68a}). Here we provide a description of
the fitting procedure; a more detailed description can be found in the
Appendix of
\citet{blanton03q}.

For each galaxy, we fit an axisymmetric \Sersic\ model of the form 
\begin{equation}
\label{sersic}
I(r) = A \exp\left[ - (r/r_0)^{1/n} \right].
\end{equation}
to the mean fluxes in annuli output by the SDSS photometric pipeline
{\tt PHOTO} in the quantities {\tt profMean} and {\tt profErr}
(\citealt{stoughton02a} list the radii of these annuli). In Equation
\ref{sersic}, $n$ is referred to as the \Sersic\ index. {\tt PHOTO}
outputs these quantities only out to the annulus which extends beyond
twice the Petrosian radius, or to the first negative value of the mean
flux, whichever is largest. In any case, we never consider data past
the 12th annulus, whose outer radius is $68.3''$. For the median
galaxy we have data and perform the fit out to $27.9''$ (the median
half-light radius from the fits is $2''$). We minimize:
\begin{equation}
\chi^2 = \sum_i [ (\mathtt{profMean}_i -
  \mathtt{sersicMean}_i(A,n,r_0) 
) / \mathtt{profErr}_i]^2
\end{equation}
where {\tt sersicMean}$_i(A,n,r_0)$ is the mean flux in annulus $i$ for
the \Sersic\ model convolved with a three-gaussian seeing model for
the given field. 

We have evaluated the performance of this algorithm in the following
way. Taking a sampling of the parameters of our fits from the Main
galaxy sample, we have generated about 1200 axisymmetric fake galaxy
images following Equation \ref{sersic} exactly, which we refer to as
``fake stamps.''  In order to simulate the performance of {\tt PHOTO},
we have distributed the fake stamps among SDSS fields. For each band,
we convert the fake stamps to SDSS raw data units, convolve with the
estimated seeing from the photometric pipelines, and add Poisson noise
using the estimates of the gain. We add the resulting image to the
SDSS raw data at a random location on the frame, including the tiny
effects of nonlinearity in the response and the less tiny flat-field
variation as a function of column on the chip. We run {\tt PHOTO} on
the resulting set of images to extract and measure objects and then
run our
\Sersic\ fitting code. This procedure thus includes the effects of
seeing, noise, and sky subtraction. We have tested that our results
remain the same if we insert images using an alternative estimate of
the seeing based simply on stacking nearby stars (still fitting using
our three-gaussian fit to the PSP seeing estimate).

\begin{figure}
\figurenum{\fignum}
\plotone{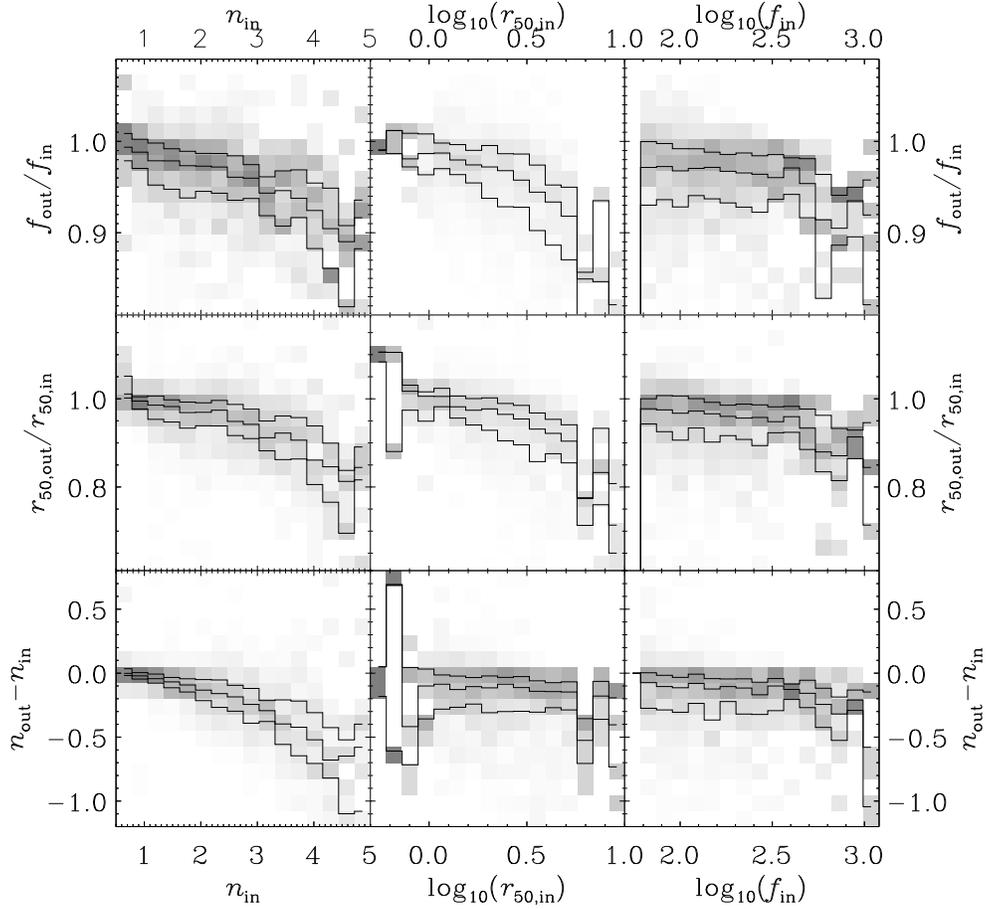}
\epsscale{1.0}
\caption{\label{sersic_compare} Residuals of the fit \Sersic\
  parameters as a function of the input \Sersic\ parameters for a set
  of 1200 simulated galaxies inserted into raw data and processed with
  the SDSS photometric pipeline plus the \Sersic\ fitting procedure.
  The fluxes and sizes are those associated with the
  \Sersic\ fit. The greyscale represents the conditional probability
  of the $y$-axis measurement given the $x$-axis input; the lines show
  the quartiles of that distribution.}
\end{figure}

Figure \ref{sersic_compare} displays the distribution of fit
parameters in the $r$-band (converting $A$ and $r_0$ to total flux $f$
and half-light radius of the profile $r_{50}$), as a function of the
input parameters. Each panel shows the conditional distribution of the
quantity on the $y$-axis as a function of quantity on the
$x$-axis. The fluxes $f$ are expressed in nanomaggies, such that
$f=100$ corresponds to $m = 17.5$, near the flux limit of the Main
galaxy sample.  The lines show the quartiles of the distribution. At
all \Sersic\ indices, sizes, and fluxes, the performance is good.

For larger sizes, sizes and fluxes are underestimated by about 10\%
and 15\% respectively, while the \Sersic\ index is constant over a
factor of ten in size. For high \Sersic\ indices, the sizes and fluxes
are slightly underestimated (again by about 10\% and 15\%) while the
\Sersic\ index itself is underestimated by (typically) -0.5 for a de
Vaucouleurs galaxy --- meaning that a true de Vaucouleurs ($n=4$)
galaxy yields $n\sim 3.5$ in our fits.  This remaining bias is not
much larger than the uncertainty itself and is comparable to the bias
one expects (in the opposite direction) from neglecting
non-axisymmetry.

The bias is partly due to our approximate treatment of the seeing, but
mostly due to small errors in the local sky level (at the level of 1\%
or less of the sky surface brightness) determined by the photometric
software. If one fits for the sky level, one can recover the \Sersic\
indices (and fluxes and sizes) of the fake galaxies far more
accurately. However, because the \Sersic\ model is not a perfect model
for galaxy profiles, for real data the fits apply unrealistically high
changes to the sky level to attain slight decreases in $\chi^2$. The
resulting sizes and fluxes of the largest and brightest galaxies are
obviously wrong. Thus, we satisfy ourselves that the measurements we
obtain with the fixed sky level yield approximately the right answer
for galaxies which are actually \Sersic\ shaped, and for other
galaxies merely supply a seeing-corrected estimate of size and
concentration.

\subsection{Distances to low redshift galaxies}
\label{distance}

At very low redshift we must take care in using the redshift as an
estimate of the distance. First, we convert the heliocentric redshift
provided by {\tt idlspec2d} into the frame of the Local Group
barycenter using the Local Group heliocentric velocity determination
of \citet{yahil77a}.

Then, we use a model of the local velocity field based on the IRAS 1.2
Jy redshift survey
determined by 
\citet{willick97a} (using $\beta = 0.5$)
in order to find the most likely distance of
the given galaxy. Along the sightline ${\hat x}$ to each galaxy we
maximize the likelihood density expressed by:
\begin{equation}
\label{likeden}
p(r) = 
r^2 \exp\left[ (cz_{\mathrm{LG}} - v(r, {\hat x}) - 
r)^2/2\sigma_v^2 \right], 
\end{equation}
where $v(r, {\hat x})$ is the outward radial peculiar velocity at
distance $r$ (expressed in km s$^{-1}$) in direction ${\hat x}$. The
fit of \citet{willick97a} extends to 64 $h^{-1}$ Mpc. Outside that
radius we neglect peculiar velocities and assume the Hubble Law is
exact. We set $\sigma_v = 150$ km s$^{-1}$ independent of local
density. We taper the peculiar velocities $v(r, {\hat x})$ to zero
between 50 and 64 $h^{-1}$ Mpc in order to provide a smooth transition
between these two regimes. The typical corrections are of the order of
200--300 km s$^{-1}$.

We report errors in the distance based on the following calculation:
we find the furthest distances above and below the best-fit distance
at which the probability in Equation \ref{likeden} is equal to
$\exp(-2)$ of its peak value (the $2\sigma$ point in Gaussian
statistics) and report $1/4$ of the difference as the standard
deviation in the distance. Near the edges of the volume for which we
have an estimate of the velocity field ($r=0$ and $r=6,400$ km
s$^{-1}$) we use $1/2$ of the difference between the best fit distance
and the inner distance satisfying the above criterion. Outside that
volume, we simply use the velocity dispersion $\sigma_v = 150$ km
s$^{-1}$.

In order to test our method, we have compared our version of distances
to a set of galaxies in common with the Mark III catalog of
\citet{willick97a}. Our results are consistent with the IRAS-predicted
velocity field distances ({\tt dist\_iras}) in that catalog. On the
other hand, there are significant disagreements (at the few Mpc level)
with the Tully-Fisher corrected distances ({\tt dist\_tfc}) of that
catalog, in the sense that the Tully-Fisher distances tend to be
higher. These differences reflect the inability of the IRAS density
field to predict velocities perfectly in the directions probed by our
sample.

The results of the procedure are in the file:
\begin{quote}
{\tt \$VAGC\_REDUX/velmod\_distance/distance\_sigv150.fits}
\end{quote}
which lists the coordinates, heliocentric ({\tt ZACT}), Local Group
relative ({\tt ZLG}), and peculiar-velocity corrected ({\tt ZDIST})
redshifts for each object.

\subsection{Matching spectra for badly deblended targets}
\label{matchspec}

When we matched the SDSS tiling catalog ({\tt
sdss\_tiling\_catalog.fits}) and the SDSS spectroscopic catalog ({\tt
sdss\_spectro\_catalog.fits}) to the SDSS imaging catalog data, we
used a match length of 2$''$. However, for low surface brightness or
complex galaxies, the behavior of the deblender has changed as the
photometric software has changed over time. Thus, the spectrum may
reflect that of an object in the latest SDSS imaging catalog, but not
be near the nominal center of that object. We would like to make sure
that we can recover the redshifts of objects in these cases.

In order to do so, we take all objects in the {\tt
object\_sdss\_imaging} file that have no spectroscopic matches, and
compare them to all spectra in the SDSS spectroscopic catalog that are
not already matched to {\tt object\_sdss\_imaging} galaxies and which
are within $2r_{P, 90}$ (twice the Petrosian 90\% light radius) of the
center of the object.  There are about 3000 spectra with such a
candidate match. For each object we take a 3$''$ radius aperture
around the center of each nearby spectrum, and measure the flux
contributed by the object in question according to the deblender (or
its parent, if the quality flag {\tt USE\_PARENT} is set) as well as
the total flux in that exact same aperture. If the flux in that
aperture from the object is at least half of the total flux in that
aperture then we consider the given spectrum to match the given
object. We perform the same operation for the tiling catalog in order
to match it to the {\tt object\_sdss\_imaging} catalog.

We store the results in the files:
\begin{quote}
{\tt \$VAGC\_REDUX/matchspec/matchspec.fits} \\
{\tt \$VAGC\_REDUX/matchspec/matchtiled.fits} 
\end{quote}
Each file contains an entry for the closest spectrum within
$2r_{P,90}$ to the center of each unmatched object (the column {\tt
OBJECT\_POSITION} indicates which object is under consideration, the
column {\tt ISP} or {\tt ITI} indicates which entry of the {\tt
sdss\_spectro\_catalog} or {\tt sdss\_tiling\_catalog}). The column
{\tt SPMATCHED} or {\tt TIMATCHED} indicates whether the entry
satisfies the criterion above. The catalog entries from {\tt
  sdss\_spectro\_catalog} or {\tt sdss\_tiling\_catalog} are also
included for convenience.

The results of this operation are included when we build the LSS
sample (Section \ref{lss}), the low-redshift catalog (Section
\ref{lowz}), the $K$-corrections (Section \ref{kcorrect}), and the
collision corrections (Section \ref{collisions}).

\subsection{Double stars}
\label{doublestar}

At very low redshifts, many of the photometrically defined
``galaxies'' are not galaxies at all, but instead are double stars
which the photometric software failed to deblend. Typically these
double stars are flagged as galaxies by the photometric pipelines,
because they are resolved, are bright ($m_r < 16$) and small
$\theta_{50} < 2''$.

In order to find many of these double stars, we have post-processed
the atlas images for all objects that the photometric software reports
as resolved and that have $cz < 1500$ km/s as well as all galaxies
without spectra with for which $\mu_{50,r} < 19$. We fit a double PSF
model to the $r$-band image using an {\tt idlutils}\footnote{\tt
http://spectro.princeton.edu} utility we wrote for this purpose called
{\tt multi\_psf\_fit}. We classify as double stars all objects that
pass both the following criteria:
\begin{eqnarray}
\frac{f_{\mathrm{model}}}{f_{\mathrm{PSF}}} &<& 2 \\
\frac{\chi_{\mathrm{double}}^2}{\chi_{\mathrm{single}}^2} &<& 0.2 ,
\end{eqnarray}
where $f_{\mathrm{model}}$ and $f_{\mathrm{PSF}}$ are the ``model''
and ``PSF'' fluxes reported by the photometric pipeline
(\citealt{abazajian04a}), and $\chi_{\mathrm{double}}^2$ and
$\chi_{\mathrm{single}}^2$ are measures from {\tt multi\_psf\_fit} of
the residuals between a model and the image using the double star and
single star models. The first criterion is necessary to exclude cases
where the galaxy model fit is much more appropriate; the second
criterion simply measures how much better the double star fit is. We
have set the parameters conservatively in the sense that there are
essentially no objects reported as double stars that are not in fact
double stars. On the other hand, this conservatism means that there
are some double stars not flagged as such by this procedure.

The results of this procedure exist in the file:
\begin{quote}
{\tt \$VAGC\_REDUX/doublestar/doublestar.fits}
\end{quote}
The important column in this file is {\tt ISDOUBLE}, which has a ``1''
if the procedure above concludes that the object is a double star, and
``0'' otherwise.

It is worthwhile noting that because M stars have such distinctive
spectra, {\it any} resolved object flagged in {\tt
object\_sdss\_spectro} as an M star is in fact a double star or an M
star in the foreground of a distant galaxy. So when searching for very
low redshift objects, one should exclude any object with a {\tt
subclass} of M star. On the other hand, it is common for low redshift
galaxies to be classified as stars of other types.

\subsection{Eyeball quality checks}
\label{eyeball}

For various purposes, we have performed a number of eyeball checks on
the photometry and the spectra. We do not claim any completeness in
terms of what set of objects we have checked, but it is usually
productive to exclude objects we have flagged as errors in this list.

The quality file is at:
\begin{quote}
{\tt \$VAGC\_REDUX/eyeball/quality.fits}
\end{quote}

For each object which we have quality checked, we have set values in a
bitmask flag whose values and meanings are listed in Table
\ref{quality}. 

For objects with {\tt DONE} set and no other flags, we have concluded
that the object is dealt with by {\tt PHOTO} more-or-less
correctly. Any other flags set indicates an error and the object
should be ignored, except in the cases that {\tt USE\_ANYWAY} is set,
which means we have concluded that keeping the object and its
measurements is better than excluding it, or that {\tt USE\_PARENT} is
set, which means we recommend using the measurement of the parent.

In cases of {\tt USE\_PARENT}, we have created a set of files for the
measurement of the parents, which is at
\begin{quote}
{\tt \$VAGC\_REDUX/parents}
\end{quote}
This directory is more-or-less constructed to be parallel with the
{\tt \$VAGC\_REDUX} directory. The list of objects 
whose parents we have processed is in the file:
\begin{quote}
{\tt \$VAGC\_REDUX/parents/object\_sdss\_imaging\_parents.fits}, 
\end{quote}
which also has the photometric information for each object (as in the
{\tt object\_sdss\_imaging} file).  We calculate $K$-corrections and
\Sersic\ fits for the parents and store them in the appropriate
directories below this level, as fully documentaed on the web site
listed in Table \ref{webpages}.

In some cases, the parent is centered in an odd place, such as an HII
region on the outskirts of the galaxy, rather than the center. These
objects we flag as {\tt BAD\_PARENT\_CENTER}. Although the redshifts
and some photometric measurements will be fine, structural
measurements (such as
\Sersic\ fits) will be misleading.

\subsection{Large-scale structure geometry}
\label{lss}

In a separate directory:
\begin{quote}
{\tt \$LSS\_REDUX/drtwo14}
\end{quote}
we store the information we use to describe the large-scale structure
geometry of the survey and with which to do large-scale structure
science. Recall that, like \vagcdir{}, \lssdir{} is an environmental
variable describing the root location of the large-scale structure
data. The relevant value will be listed in the online documentation,
as it will in principle change over time.

There are two basic files. First, there is one file fully describing the
geometry  
\begin{quote}
{\tt lss\_combmask.drtwo14.fits} 
\end{quote}
which is readable with the {\tt idlutils} routine {\tt
read\_fits\_polygons} and which contains a row for every spherical
polygon. This geometry is the area covered by the imaging, by the set
of survey tiles, and not near Tycho stars (as described below).
\begin{enumerate}
\item {\tt SECTOR}: sector number (see description of tiling geometry
  in Section \ref{tiling})
\item {\tt FGOTMAIN}: fraction of Main sample targets which have good
redshifts in this sector
\item {\tt MMAX}: flux limit based on targeting limit and change in
  calibration since targeting ($r$-band magnitudes)
\item {\tt DIFFRUN}: whether the ``best'' imaging in this field is the
  same observation as the ``target'' imaging 
\item {\tt ITILING\_GEOMETRY}: position (zero-indexed) of the tiling
  polygon (see Section \ref{tiling}) to which this polygon belongs
\item {\tt ITARGET\_GEOMETRY}: position (zero-indexed) of the target
  polygon to which this polygon belongs (in the file \vagcdir{\tt
  /sdss/sdss\_target\_geometry.fits}) 
\item {\tt ILSS}: index into the {\tt lss\_geometry} file described
below
\item {\tt RA}: estimated center of the field this was based on (J2000
degrees)
\item {\tt DEC}: estimated center of the field this was based on (J2000
degrees)
\end{enumerate}
{\tt RA} and {\tt DEC} are not always inside the given polygon;
however, the polygon is {\it always} fully contained within a 0.36 deg
circle surrounding that center. 

Note that {\tt MMAX} varies with position for two reasons. First, the
explicit targeting limits changed with time over the course of the
survey. Second, the calibration of the data has improved since the
targeting, such that the flux limit in the recalibrated data varies
across the sky. {\tt MMAX} accounts for both effects.

There is an IDL utility in the {\tt vagc} product called {\tt
get\_sdss\_icomb} which takes right ascension and declination values,
and quickly checks which row of the {\tt lss\_combmask} file it is
contained in. Because this piece of code is relatively simple and
self-contained, we reproduce it in full in the Appendix.

Second, there is a  file describing the relationship each object in
{\tt object\_sdss\_imaging} has with the geometry: 
\begin{quote}
{\tt lss\_index.drtwo14.fits} 
\end{quote}
This file contains the columns:
\begin{enumerate}
\item {\tt RA}: right ascension (J2000) in degrees
\item {\tt DEC}: declination (J2000) in degrees
\item {\tt Z}: redshift (from SDSS, 2dFGRS, PSC$z$, or RC3, in that
  order of preference)
\item {\tt ZTYPE}: which catalog (one-indexed) from the above list the
  redshift comes from (zero if none)
\item {\tt SECTOR}: which sector the object is in
\item {\tt FGOTMAIN}: Main sample completeness for each object (as
described above)
\item {\tt MMAX}: what the nominal flux limit is in the area of sky
  this object is in
\item {\tt ILSS}: which polygon in the {\tt lss\_geometry} file this
  object is in (-1 if none)
\item {\tt ICOMB}: which polygon in the {\tt lss\_combmask} file this
  object is in (-1 if none)
\item {\tt VAGC\_SELECT}: how this object was selected (see Section
  \ref{sdss_imaging}) 
\end{enumerate}
This file can be used to make some simple cuts to select galaxies with
redshifts. The SDSS redshifts include those which have the {\tt
ZWARNING} flag set but we have determined to be good and flagged {\tt
GOOD\_Z} (see Section \ref{eyeball}), as well as those which we have
matched using the extended matching criterion described in Section
\ref{matchspec}.

In addition, there are two files describing the geometry, out of which
we have built {\tt lss\_combmask}:
\begin{quote}
{\tt lss\_geometry.drtwo14.fits} \\
{\tt lss\_bsmask.drtwo14.fits} 
\end{quote}
Each of these files contain sets of spherical polygons. The first file
contains the geometry of the survey as a whole.  The second file
contains a mask cut out around bright stars (stars with $B<13$ in the
Tycho catalog; \citealt{hog00a}). The radius of the circle around each
bright star is set according to the formula:
\begin{equation}
\theta = (0.0802 B'^2 - 1.86 B' +11.625)
\end{equation}
where $\theta$ is in arcmin, $B'$ is the Tycho magnitude $B$ if $6 < B
< 11.5$, but $B'=6$ if $B<6$ and $B'=11.5$ if $B>11.5$. This radius is
about that at which the mean density of galaxies near Tycho stars
drops to half the background value (I.~Strateva, private
communication). The mask is designed such that each polygon describing
it is fully contained in a single polygon of the {\tt lss\_geometry}
file (which one is stored in the {\tt ILSS} column).

Finally, the directory:
\begin{quote}
{\tt \$LSS\_REDUX/drtwo14/random/}
\end{quote}
contains one hundred random catalogs with one million points each,
distributed with constant surface density in the area included by {\tt
lss\_combmask}. In addition to right ascension and declination, this
file contains {\tt ILSS}, indicating which polygon of {\tt
lss\_geometry} the random point is in, and {\tt EBV}, the $E(B-V)$
value for this direction from
\citet{schlegel98a}.   

\section{A low redshift catalog ($0.0033 < z < 0.05$)} 
\label{lowz}

One of the areas of the SDSS which requires special care is in the
treatment of galaxies at low redshift. In order to study the property
of galaxies at low redshifts and, correspondingly, at low
luminosities, we have done some simple checks of the SDSS catalog in
this regime, cleaned up the catalog where it was simple to do so, and
put together a ``low-redshift'' catalog of galaxies with estimated
comoving distances in the range $10 < d < 150$ $h^{-1}$ Mpc. 

For the purposes of this catalog, we checked the atlas images and
spectra of a number of galaxies.  We flagged bad deblends as errors
and set other quality flags according to Section \ref{eyeball}.  In
particular, in DR2 we have checked all objects in the catalog that we
have selected as Main-sample-like objects (see Section
\ref{sdss_imaging}), that have a spectrum, and that satisfy one of the
following criteria: 
\begin{enumerate}
\item $M_r > -15$ and $z>0.003$, if a good redshift exists in the
  sample redshift range, the object is not classified as a double star
  according the algorithm in Section \ref{doublestar}, and it is not
  classified as an M star. We deemed about 22\% of these to be
  deblending errors in the latest reductions; for about 72\% of these
  errors (16\% of the total number) using the parent photometry is
  sufficient. So we recover about 94\% of the objects in this
  category.
\item $0.003 < z < 0.01$, if a good redshift exists, the object
  is not classified as a double star according the algorithm in
  Section \ref{doublestar}, and it is not a star. Again, about 19\%
  are deblending errors, for 70\% of which (14\% of the total number)
  the parent photometry is sufficient, thus recovering about 96\% of
  the galaxies in this category.
\item The spectroscopic pipeline warning flag {\tt ZWARNING} is set to
  something non-zero, indicating a problem. For many of these (about
  25\%) the software in fact finds a correct redshift, and we flag
  these as {\tt GOOD\_Z} (see Table \ref{quality}). This yields 97
  more galaxies in our sample.
\item The galaxy is in a pair with a separation less than 50 $h^{-1}$
  kpc and an angular separation less than 0.1 deg. About 12\% of these
  are incorrectly deblended in this sample, about 37\% of which (4\%
  of the full sample) could be replaced with the parent photometry,
  yielding a recovery rate of 92\% of the galaxies in this category.
\end{enumerate}
Thus, after performing these checks, we are correctly treating the
photometry for $>90\%$ of the galaxies for which we have spectroscopy.

We select galaxies from the NYU-VAGC for the low-redshift sample
according to the following criteria:
\begin{enumerate}
\item The galaxy is selected by the Main sample-like criterion defined
in Section \ref{sdss_imaging} (that is, the bitmask {\tt
VAGC\_SELECT} has bit 2 set).
\item The quality bit has no flags set except for {\tt DONE} and/or
{\tt GOOD\_Z} or {\tt BAD\_PARENT\_CENTER}; {\it or} it has {\tt
USE\_ANYWAY} or {\tt USE\_PARENT} set.
\item The spectroscopic subclass does not indicate an M star. 
\item We have not classified the object as a double star (see Section
\ref{doublestar}). 
\item There is a good SDSS redshift determination 
(meaning the spectroscopic flag {\tt ZWARNING} was zero or the
  quality flag {\tt GOOD\_Z} was non-zero).
\item The distance estimate from Section \ref{distance} yields a
distance redshift within the range $0.0033 < z < 0.05$ (see below).
\item The object is within the large-scale structure geometry and
outside the bright star mask (see Section \ref{lss}).
\item The Galactic extinction corrected, $r$ band Petrosian magnitude
is less than the flux limit in the given direction (see Section
\ref{lss}). 
\end{enumerate}
The redshifts we use include those matched to the imaging objects
using the criterion described in Section \ref{matchspec}.  Note that
we do not require that the object be spectroscopically classified as a
galaxy; a certain number of low redshift galaxies are classified as
stars, especially in cases that the spectrum has a low signal-to-noise
ratio.  The low redshift cut corresponds to $10$ $h^{-1}$ Mpc. We do
not consider anything below this redshift because the sample becomes
highly incomplete (due to shredding by the photometric pipeline of
large resolved galaxies) and the distance estimates for such objects
are highly affected by peculiar velocities.

For galaxies with {\tt USE\_PARENT} set, we replace the SDSS child's
photometry with that of the parent, using the results described in
Section \ref{eyeball}. 

In the DR2 area, 28,089 galaxies pass the above criteria. Weighted by
the completeness, the effective area of the sample is 2220.9 square
degrees.

To compute the global number densities of galaxies as a function of
their properties, it is necessary to compute the number-density
contribution $1/\Vmax$ for each galaxy, where $\Vmax$ is the volume
covered by the survey in which this galaxy could have been observed,
accounting for the flux, redshift limits, and completeness as a
function of angle (Schmidt 1968). We calculate this volume as follows:
\begin{equation}
\Vmax = \int\mathrm{d}\Omega\,f(\theta,\phi)\, 
\int_{z_{\mmin}}^{z_{\mmax}(\theta,\phi)}
\mathrm{d}z\,\frac{dV}{dz} \;,
\end{equation}
where $f(\theta,\phi)$ is the spectroscopic completeness as a function
of angle, averaging about 90\% across the survey. We determine this on
a sector-by-sector basis (it is the {\tt FGOTMAIN} value described in
Section
\ref{lss}). $z_{\mmax}(\theta,\phi)$ is defined for this sample by:
\begin{eqnarray}\displaystyle
z_{\mmax}(\theta,\phi) &=& \mmin(
z_{m,\mmax}(\theta,\phi), 
0.05) \;, \cr
z_{\mmin} &=& 0.0033 \; .
\end{eqnarray}
The flux limit $m_{r,\mmax}(\theta,\phi)$ (usually 17.77~mag)
implicitly sets $z_{m,\minmax}(\theta,\phi)$ by:
\begin{eqnarray}\displaystyle
m_{r,\mmax}(\theta,\phi) &=& M_r +
\mathrm{DM}(z_{m,\mmax}(\theta,\phi))\cr
& & + K_r(z_{m,\mmax}(\theta,\phi)) \;.
\end{eqnarray}
Note that over this redshift range we can ignore the contribution of
the surface brightness limits to $\Vmax$, since cosmological surface
brightness dimming is such a small effect. 

There is a complication at low redshift: our estimate of
the luminosity has a large uncertainty due to galaxy peculiar
velocities. \Vmax\ as calculated above is a nonlinear function of
luminosity, such that an uncertainty in luminosity will yield an {\it
underestimate} of \Vmax. For a fair estimate of \Vmax, we use:
\begin{equation} 
\Vmax = \int dL p(L) \Vmax(L)
\end{equation} 
where $p(L)dL = p(r) dr$ (from Equation \ref{likeden} above). This
estimate is an average of \Vmax\ for all possible luminosities based on
the probability of the galaxy having that luminosity. In practice,
there is only a small difference between the results one finds using
the regular $\Vmax$ estimator and this one.




The catalog is available in the file:
\begin{quote}
{\tt \$VAGC\_REDUX/lowz/lowz\_catalog.drtwo14.fits}
\end{quote}
Its columns are described in Table \ref{lowz_catalog}. 


Atlas images (that is, images with neighboring objects removed;
\citealt{stoughton02a}) of all the
objects in the catalog are available in the directories:
\begin{quote}
{\tt \$VAGC\_REDUX/lowz/images/00h}\\
{\tt \$VAGC\_REDUX/lowz/images/01h}\\
\ldots \\
{\tt \$VAGC\_REDUX/lowz/images/23h}
\end{quote}
where each directory contains galaxies in a particular hour of right
ascension (J2000). The names of each atlas image are based on the IAU
name of each object; eg.
\begin{quote}
{\tt lowz-atlas-J044112.00+003202.3.fits}
\end{quote}
As described on the web site, each file contains ten HDUs. HDUs 0, 2,
4, 6, and 8 contain the $ugriz$ images of each galaxy. HDUs 1, 3, 5,
7, and 9 contain the inverse variance $ugriz$ images of each
galaxy. In addition, there is a file of the form:
\begin{quote}
{\tt psf-J044112.00+003202.3.fits}
\end{quote}
which contains five HDUs, the estimated PSFs at the center of the
object in $ugriz$. 

We have already used this low redshift catalog for an investigation of
the population of low luminosity galaxies in the SDSS survey (Blanton
et al. in preparation). We expect the catalog and images to be useful
in a number of other ways. For example, it provides a nice low
redshift sample for comparison to high redshift galaxy samples.

\section{Software tools}
\label{tools}

Generally speaking we have used a combination of C and IDL in
constructing this catalog. Astronomers will likely find many of the
utilities we have used to construct the dataset useful for analyzing
it as well. The source code and online documentation for some of these
tools is listed in Table \ref{webpages}.

Here are some short descriptions of the tools themselves:
\begin{enumerate}
\item {\tt idlutils}: A general astronomically useful set of IDL
utilities maintained by David Schlegel and Doug Finkbeiner,
incorporating the Goddard IDL library maintained by Wayne Landsman,
and contributed to by many others too numerous to mention. In
particular, this library contains readers for the FITS files,
spherical polygon files (see Section \ref{window} below), and FTCL
parameter files (a special ASCII format) that our catalog contains. In
particular:
\begin{enumerate}
\item {\tt mrdfits} is a general FITS reader, which can read FITS
images and tables ({\tt .fits} files)
\item {\tt read\_mangle\_polygons} will read spherical polygon files
produced by {\tt mangle} ({\tt .ply} files) and {\tt
read\_fits\_polygons} will read spherical polygons files in FITS format
(and {\tt idlutils} has code which makes it easy to work with the
resulting structures). See the {\tt mangle} web site listed in Table
\ref{webpages}, or Section \ref{window} below, for details on spherical
polygon files. 
\item {\tt yanny\_readone} is a FTCL parameter file reader ({\tt .par}
files, a special SDSS ASCII format; see the DR2 web site listed in
Table \ref{webpages} for details)
\end{enumerate}
\item {\tt photoop}: An SDSS-specific set of utilities written in
Perl, IDL, and C by David Schlegel, Doug Finkbeiner, and Nikhil
Padmanabhan. These tools are primarily for performing
the photometric reductions and calibrations, but also contain image
analysis utilities that may be useful to users.
\item {\tt vagc}: The code responsible for producing this catalog. It
contains some useful utilities for reading in the data from the
catalog.
\item {\tt kcorrect}: A set of utilities for calculating
$K$-corrections and photometric redshifts, tuned to work for SDSS and
2MASS data (\citealt{blanton03b}). Our catalog contains
$K$-corrections already calculated, but the user may find this useful.
\item {\tt mangle}: A general set of tools for handling window
functions on the sphere, developed by \citet{hamilton03a}. It is
described more fully in Section \ref{window}. These routines can be
useful to the user for creating random catalogs from the geometrical
descriptions given here, and also for checking whether certain
directions are inside or outside the surveys. This code is also
distributed as part of {\tt idlutils}.
\end{enumerate}

We should note that {\it none} of these tools are {\it necessary} for
using the data, only recommended as generally useful packages.

\section{Summary}
\label{summary}

Here we have presented a catalog of galaxies combining information
from SDSS, 2MASS, FIRST, PSC$z$, RC3, and 2dFGRS. The main
improvements of this catalog over the standard SDSS release (in
addition to the matches to other catalogs) are a better calibration
and an explicit description of the geometry, including all of the
information necessary to perform large-scale structure analyses with
the catalog. We have also included structural measurements,
$K$-corrections, peculiar velocity corrections, and quality checks of
many objects. The catalog is fully documented on the web site listed
in Table \ref{webpages}.

As we continue to develop the NYU-VAGC, we expect to add considerable
functionality. For example:
\begin{enumerate}
\item We plan to extract images and, where possible,  measure
structural parameters for galaxies detected in other surveys which
overlap the SDSS. 
\item We plan to include parameters for NYU-VAGC galaxies from other
surveys ({\it e.g.} the Spitzer Space Telescope and the Galaxy
Evolution Explorer).
\item We plan to continue adding SDSS data as it is released, and
improving the treatment of deblended galaxies.
\item The SDSS ubercalibration procedure will continue to improve, as
the SDSS collaboration takes more and more overlapping data.
\end{enumerate}

We note that this project will be more successful if users provide
feedback about how the catalog could be improved, since we do not
expect that we can predict from pure thought what astronomers will
find useful.

\acknowledgments

MB and DWH acknowledge NASA NAG5-11669 for partial support.  We thank
Mulin Ding, Steve Huston, and Craig Loomis for taking excellent care
of the computing systems used for this project. We thank Iskra
Strateva for sharing her work on the effects of bright stars on galaxy
detection in the SDSS. We thank Andrew Hamilton for early access to
his {\tt mangle} utilities and for his tremendous help in using them.
We thank the 2dFGRS team, the PSC$z$ team, and the FIRST team for the
public release of their catalogs.

The creation and distribution of the NYU-VAGC is funded by the New
York University Department of Physics.  We are greatly indebted to the
SDSS team as a whole, in the form of its Builders, its Participants,
and its External Participants, as well as the investment made by its
sponsors.
Funding for the creation and distribution of the SDSS Archive is
provided by the Alfred P. Sloan Foundation, the Participating
Institutions, the National Aeronautics and Space Administration, the
National Science Foundation, the U.S. Department of Energy, the
Japanese Monbukagakusho, and the Max Planck Society. The SDSS Web site
is {\tt http://www.sdss.org/}.

The SDSS is managed by the Astrophysical Research Consortium (ARC) for
the Participating Institutions. The Participating Institutions are The
University of Chicago, Fermilab, the Institute for Advanced Study, the
Japan Participation Group, The Johns Hopkins University, Los Alamos
National Laboratory, the Max-Planck-Institute for Astronomy (MPIA),
the Max-Planck-Institute for Astrophysics (MPA), New Mexico State
University, University of Pittsburgh, Princeton University, the United
States Naval Observatory, and the University of Washington.

This publication makes use of data products from the Two Micron All
Sky Survey, which is a joint project of the University of
Massachusetts and the Infrared Processing and Analysis
Center/California Institute of Technology, funded by the National
Aeronautics and Space Administration and the National Science
Foundation.

This research has made use of the NASA/IPAC Extragalactic Database
(NED) which is operated by the Jet Propulsion Laboratory, California
Institute of Technology, under contract with the National Aeronautics
and Space Administration.

\appendix

\section*{Retrieving completeness and flux limit given a direction}

Here we reproduce the IDL utility from the {\tt vagc} product called
{\tt get\_sdss\_icomb} described in Section \ref{lss}.  The routine
depends on having the {\tt idlutils} software installed.  It takes
right ascension and declination values, and quickly checks which row
of the {\tt lss\_combmask} file refers to that direction. The
appropriate row of that file has information on the completeness and
flux limit in that direction. 

\begin{verbatim}
;+
; NAME:
;   get_sdss_icomb
; PURPOSE:
;   return indices of polygons in lss_combmask for given ra, dec list
; CALLING SEQUENCE:
;   icomb= get_sdss_icomb(ra, dec, sample=sample)
; INPUTS:
;   ra - [N] right ascension, degrees
;   dec - [N] declination, degrees
;   sample - name of LSS sample to use
; OPTIONAL INPUTS/OUTPUTS:
;   combmask - combined mask file
; OUTPUTS:
;   icomb - [N] index of each ra and dec into lss_geometry
; COMMENTS:
;   reads the $LSS_REDUX/sample/lss_combmask.sample.fits file
;   ra, dec locations outside the geometry are assigned icomb=-1
;   use the combmask input/output keyword to save combmask for
;     multiple calls 
; REVISION HISTORY:
;   28-Mar-2004  Written by Mike Blanton, NYU
;-
;------------------------------------------------------------------------------
function get_sdss_icomb, ra, dec, sample=sample, combmask=lss_combmask

if(n_tags(lss_combmask) eq 0) then $
  read_fits_polygons, $
  getenv('LSS_REDUX')+'/'+sample+'/lss_combmask.'+sample+'.fits', lss_combmask

icomb=lonarr(n_elements(ra))-1L
spherematch,ra,dec,lss_combmask.ra, lss_combmask.dec,0.36,m2,m1,d12,maxmatch=0
im1sort=sort(m1)
im1uniq=uniq(m1[im1sort])
istart=0
for i=0L, n_elements(im1uniq)-1L do begin
    iend=im1uniq[i]
    icurr=im1sort[istart:iend]
    icurrlss=m1[icurr[0]]
    ipossible=m2[icurr]
    indxpoly= $
      where(is_in_polygon(ra=ra[ipossible],dec=dec[ipossible], $
                          lss_combmask[icurrlss]), countpoly)
    if(countpoly gt 0) then $
      icomb[ipossible[indxpoly]]=icurrlss
    istart=iend+1L
endfor

return, icomb

end
\end{verbatim} 

\newpage

\clearpage
\clearpage
\begin{landscape}
\begin{deluxetable}{llll}
\tablewidth{0pt}
\tablecolumns{4}
\tablecaption{\label{webpages} Related web pages}
\tablehead{ Survey & Description & Tools & URL }
\startdata
SDSS & DR2 data release & & {\tt http://www.sdss.org/dr2} \cr
 & Value-added catalog & {\tt vagc} & {\tt http://sdss.physics.nyu.edu/vagc} \cr
 & Princeton-MIT spectroscopic reductions & {\tt idlutils} & {\tt
http://spectro.princeton.edu} \cr
 & Princeton-NYU photometric reductions & {\tt photoop} & {\tt
http://photo.astro.princeton.edu} \cr
 & $K$-corrections & {\tt kcorrect} & {\tt
http://sdss.physics.nyu.edu/vagc/kcorrect.html} \cr
2MASS & Final data release & & {\tt http://www.ipac.caltech.edu/2mass}
\cr
2dFGRS & Final data release & & {\tt
http://msowww.anu.edu.au/2dFGRS/Public/Release} \cr
PSC$z$ & Final data release & & {\tt
http://www-astro.physics.ox.ac.uk/\~{ }wjs/pscz\_data.html} \cr
RC3 & Version v3.9b & & {\tt
http://spider.ipac.caltech.edu/staff/hgcjr/rc3}
\cr
ZCAT & As of November 2003 & & {\tt
http://cfa-www.harvard.edu/\~{ }huchra/zcat} \cr
--- & Spherical polygon code & {\tt mangle} & {\tt
http://casa.colorado.edu/\~{ }ajsh/mangle} \cr
\enddata
\end{deluxetable}

\clearpage
\begin{deluxetable}{lll}
\tablewidth{0pt}
\tablecolumns{3}
\tablecaption{\label{quality} Eyeball quality flags}
\tablehead{ Bit & Name & Description }
\startdata
0 & {\tt DONE} & we have checked this object \cr
1 & {\tt OTHER} & there is something wrong which doesn't fit any of the
classes below \cr
2 & {\tt UNCLASSIFIABLE} & we can't determine whether or not something is
wrong \cr
3 & {\tt NEED\_BIGGER\_IMAGE} & we looked at too small an image
to figure out what is going on\cr
4 & {\tt BAD\_DEBLEND} & ignore anything with this
flagged unless {\tt USE\_ANYWAY} or {\tt USE\_PARENT} is set \cr
5 & {\tt FLECK} & this object has only a tiny bit of the total flux of the
full object \cr
6 & {\tt DOUBLE\_STAR} & a double star \cr
7 & {\tt HII} & an HII region plucked out of a larger galaxy \cr
8 & {\tt USE\_ANYWAY} & there is a bad deblend, but we recommend you
include the object in your sample anyway \cr
9 & {\tt EDGE} & something is wrong due to the object encountering an
edge of a field \cr
10 & {\tt SATELLITE} & this object is actually a satellite flying
overhead \cr
11 & {\tt PLANE} & this object is actually a plane flying overhead\cr
12 & {\tt BAD\_Z} & the listed redshift is wrong and 
the spectro outputs think it is right\cr
13 & {\tt INTERNAL\_REFLECTION} & strongly
affected by reflection inside the instrument \cr
14 & {\tt BAD\_SPEC\_CLASS} & the spectroscopic classification is wrong
\cr
15 & {\tt USE\_PARENT} & we recommend the parent object of this
bad deblend (see {\tt BAD\_PARENT\_CENTER} below) \cr
16 & {\tt IN\_HUGE\_OBJECT} & even the parent of this object is part of
something even bigger. \cr
17 & {\tt STAR\_ON\_GALAXY} & star sitting in
the foreground of a distant galaxy (resulting in ``galaxy''
with very small redshift) \cr
18 & {\tt QSO\_ON\_GALAXY} & QSO in the
background of a galaxy (resulting very high luminosity ``galaxy'')
\cr
19 & {\tt NEGATIVE\_QSO\_FIT} & usually indicative of a redshift
error. \cr
20 & {\tt BAD\_SPECTRUM} & the spectrum itself is bad (for example,
contaminated by scattered light from a nearby bright star)\cr
21 & {\tt POSSIBLE\_LENS} & this object is possibly a multiply-imaged
lens object\cr
22 & {\tt IS\_STAR} & although classified as a resolved object, this is
definitely a star\cr
23 & {\tt DOUBLE\_Z} & the spectrum has two potential redshifts \cr
24 & {\tt PLANETARY\_NEBULA} & the spectrum is actually that of a
planetary nebula in the foreground\cr
25 & {\tt BAD\_PARENT\_CENTER} & the parent is centered in an odd
place, which will probably confuse structural measurements\cr
26 & {\tt GOOD\_Z} & the redshift reported by the SDSS spectroscopic
pipeline is good, even though the {\tt ZWARNING} value is non-zero.\cr
\enddata
\tablecomments{The names of the flags are in the {\tt idlutils} file {\tt
\$IDLUTILS\_DIR/data/sdss/sdssMaskbits.par} and can be accessed using
the IDL functions {\tt sdss\_flagname()} and {\tt sdss\_flagval()} in
{\tt idlutils} (the bitmask name is {\tt Q\_EYEBALL}).}
\end{deluxetable}
\end{landscape}

\clearpage
\begin{deluxetable}{ll}
\tablewidth{0pt}
\tablecolumns{2}
\tablecaption{\label{lowz_catalog} Parameters in low redshift catalog}
\tablehead{ Column & Description }
\startdata
{\tt OBJECT\_POSITION} & index of object in full NYU-VAGC \cr
{\tt ICOMB} & index indicating location in SDSS geometry
(see Section \ref{lss} for details) \cr
{\tt SECTOR} & sector number in SDSS survey geometry (used to
calculate completeness, see Section \ref{lss} for details) \cr
{\tt VAGC\_SELECT} & bit mask indicating how this object was
selected (see Section \ref{sdss_imaging} for details) \cr
{\tt FGOTMAIN} & completeness of SDSS Main sample in this sector
(that is, $f(\theta, \phi)$) \cr
{\tt RA} & right ascension, J2000, in degrees \cr
{\tt DEC} & declination, J2000, in degrees \cr
{\tt RUN} & SDSS drift scan number \cr
{\tt RERUN} & SDSS processing rerun  \cr
{\tt CAMCOL} & SDSS camera column \cr
{\tt FIELD} & SDSS field \cr
{\tt ID} & SDSS object id within the field \cr
{\tt OBJC\_ROWC} & CCD $y$ position in the field (centers of
pixels are half-integers) \cr
{\tt OBJC\_COLC} & CCD $x$ position in the field (centers of
pixels are half-integers) \cr
{\tt PLATE} & SDSS spectroscopic plate \cr
{\tt FIBERID} & SDSS spectroscopic fiber number \cr
{\tt MJD} & date of SDSS spectroscopic observation \cr
{\tt QUALITY} & eyeball quality flag (see Section \ref{quality}
for full description) \cr
{\tt ABSMAG[8]} & absolute magnitude (AB) in the $ugrizJHK_s$
bandpasses (first five from SDSS Petrosian \cr
& magnitude, last three 
 from 2MASS XSC converted to the AB system as described in the text)
\cr
& $K$-corrected and Galactic extinction corrected
(\citealt{schlegel98a}). \cr
{\tt ABSMAG\_IVAR[8]} & inverse variance of uncertainty in
absolute magnitude (including the effects of errors \cr
& in the distance)\cr
{\tt MU50R} & Petrosian half-light surface brightness (magnitudes
in a square arcsecond) \cr
{\tt KCORRECT[8]} & estimated $K$-corrections in $ugrizJHK_s$
(magnitudes) \cr
{\tt PETROFLUX[5]} & SDSS Petrosian flux in $ugriz$ (nanomaggies; see
Equation \ref{nmgy} in the text) \cr
{\tt PETROTHETA[5]} & SDSS Petrosian radius in $ugriz$ (arcsec) \cr
{\tt PETROTH50[5]} & SDSS Petrosian 50\% light radius in $ugriz$
(arcsec) \cr
{\tt PETROTH90[5]} & SDSS Petrosian 90\% light radius in $ugriz$
(arcsec) \cr
{\tt EXTINCTION[5]} & Galactic extinction from
\citet{schlegel98a} in $ugriz$ (magnitudes) \cr
{\tt SERSIC\_N[5]} & \Sersic\ index estimated from radial profile
in $ugriz$ \cr
{\tt SERSIC\_TH50[5]} & 50\% light radius from \Sersic\ fit 
in $ugriz$ (arcsec) \cr
{\tt SERSIC\_FLUX[5]} & total flux from \Sersic\ fit in $ugriz$
(nanomaggies; see Equation \ref{nmgy} in the text)  \cr
{\tt VDISP} & estimated velocity dispersion from spectrum \cr
{\tt VDISP\_ERR} & estimated uncertainty in velocity dispersion
from spectrum \cr
{\tt CLASS} & spectroscopic classification, as output by the SDSS
spectroscopic pipeline (note that occasionally \cr 
& this is incorrect;
in
particular, a number of galaxies in our sample are classified as \cr
& stars
spectroscopically because the signal-to-noise 
ratio of the spectrum 
does not allow\cr & reliable discrimination between the two).  \cr
{\tt SUBCLASS} & spectroscopic subclassification (e.g., stellar
type), as output by the SDSS spectroscopic pipeline\cr
{\tt VMAX} & maximum volume in the sample over which we could
have observed this object \cr
{\tt NEDNAME} & name of NED match \cr
{\tt NEDCZ} & redshift from NED match \cr
{\tt ZLG} & Local Group relative redshift from SDSS \cr
{\tt ZDIST} & peculiar velocity corrected Local Group relative
redshift from SDSS \cr
{\tt ZDIST\_ERR} & uncertainty in {\tt ZDIST} 
\enddata
\tablecomments{These are the parameters included in the catalog
described in Section \ref{lowz}.}
\end{deluxetable}

\clearpage
\clearpage

\setcounter{thefigs}{0}

\clearpage
\stepcounter{thefigs}

\clearpage
\stepcounter{thefigs}

\clearpage
\stepcounter{thefigs}

\clearpage
\stepcounter{thefigs}

\clearpage
\stepcounter{thefigs}

\clearpage
\stepcounter{thefigs}

\clearpage
\stepcounter{thefigs}

\clearpage
\stepcounter{thefigs}

\clearpage
\stepcounter{thefigs}



\end{document}